\documentclass[useAMS,usenatbib]{mnras}

\usepackage{graphicx}
\usepackage{psfig}
\usepackage{amssymb,amsmath}
\usepackage{mathrsfs}
\usepackage{times}
\usepackage{setspace}
\usepackage{color}
\usepackage{natbib}

\def\araa{ARA\&A}%
\def\apj{ApJ}%
\def\apjl{ApJ}%
\def\apjs{ApJS}%
\def\ao{Appl.~Opt.}%
\def\aap{A\&A}%
\def\mnras{MNRAS}%
\def\prc{Phys.~Rev.~C}%
\def\prd{Phys.~Rev.~D}%
\def\solphys{Sol.~Phys.}%
\def\nat{Nature}%
\def\jgr{J.~Geophys.~Res.}%
\newcommand{\lumanue}{L_{\bar{\nu}_e}}
\newcommand{\lumnue}{L_{{\nu}_e}}
\newcommand{\begit}{\begin{itemize}}
\newcommand{\enit}{\end{itemize}}
\newcommand{\begen}{\begin{enumerate}}
\newcommand{\enen}{\end{enumerate}}

\newcommand{\beq}{\begin{equation}}
\newcommand{\eeq}{\end{equation}}
\newcommand{\beqa}{\begin{eqnarray}} 
\newcommand{\eeqa}{\end{eqnarray}} 
\def\lesssim{\mathrel{\hbox{\rlap{\hbox{\lower5pt\hbox{$\sim$}}}\hbox{$<$}}}}
\def\gtrsim{\mathrel{\hbox{\rlap{\hbox{\lower5pt\hbox{$\sim$}}}\hbox{$>$}}}}

\title[Heavy element nucleosynthesis in magnetized proto-neutron star winds]
{Heavy element nucleosynthesis in high-entropy ejections from magnetized proto-neutron star winds}

\author[Thompson \& ud-Doula]{
Todd A.~Thompson$^1$ \& Asif ud-Doula$^2$\\
$^1$Department of Astronomy and Center for Cosmology \& Astro-Particle Physics,
The Ohio State University, Columbus, Ohio 43210, USA\\
$^2$Penn State Worthington Scranton, Dunmore, PA 18512, USA}

\begin{document}

\maketitle

\label{firstpage}

\begin{abstract}
Although initially thought to be promising for production of the $r$-process nuclei, standard models of neutrino-heated winds from proto-neutron stars (PNSs) do not reach the requisite neutron-to-seed ratio for production of the lanthanides and actinides. However, the abundance distribution created by the $r$, $rp$, or $\nu p$-processes in PNS winds depends sensitively on the entropy and dynamical expansion timescale of the flow, which may be strongly affected by high magnetic fields. Here, we present results from magnetohydrodynamic simulations of non-rotating neutrino-heated PNS winds with strong dipole magnetic fields from $10^{14}-10^{16}$\,G, and assess their role in altering the conditions for nucleosynthesis. The strong field forms a closed zone and helmet streamer configuration at the equator, with episodic dynamical mass ejections in toroidal plasmoids. We find dramatically enhanced entropy in these regions and conditions favorable for third-peak $r$-process nucleosynthesis if the wind is neutron-rich. If instead the wind is proton-rich, the conditions will affect the abundances from the $\nu p$-process. We quantify the distribution of ejected matter in entropy and dynamical expansion timescale, and the critical magnetic field strength required to affect the entropy. For $B\gtrsim10^{15}$\,G, we find that $\gtrsim10^{-6}$\,M$_{\odot}$ and up to $\sim10^{-5}$\,M$_{\odot}$ of high entropy material is ejected per highly-magnetized neutron star birth in the wind phase, providing a mechanism for prompt heavy element enrichment of the universe. Former binary companions identified within (magnetar-hosting) supernova remnants, the remnants themselves, and runaway stars may exhibit overabundances. We provide a comparison with a semi-analytic model of plasmoid eruption and discuss implications and extensions.
\end{abstract}


\section{Introduction}
\label{section:introduction}

After the collapse of a massive star and subsequent explosion, a cooling proto-neutron star (PNS) remains, driving a thermal neutrino-heated transonic wind into the surrounding post-shock medium that subsides on the Kelvin-Helmholz cooling timescale $t_{\rm KH}\sim10$\,s  \citep{duncan86,woosley_baron,bhf,janka_muller}. This expanding bubble was suggested as the site in Nature for production of the heavy $r$-process nuclides \citep{woosley_hoffman,meyer92,woosley94}, but models of the outflow fail to produce the conditions required for a successful third-peak nucleosynthesis \citep{takahashi94,qian_woosley}, even including General Relativistic effects \citep{cardall_fuller,otsuki,tbm,wanajo01}. Although the light $r$-process elements can be produced in abundance, the entropy, expansion timescale, and neutron fraction of model outflows yield a neutron-to-seed ratio that is simply too small for production of the heavy $r$-process nuclei. Indeed, early models artificially increased the entropy by large factors by decreasing the outflow density in order to get heavy element abundances in agreement with the scaled Solar $r$-process distribution (e.g., \citealt{takahashi94,hoffman97}). 

Recent calculations of PNS cooling and the wind epoch indicate that the outflow might be less neutron rich than previously expected over a significant fraction of $t_{\rm KH}$ \citep{fischer2010,hudepohl2010,roberts12_wind}, further circumscribing or eliminating the potential role of normal PNS winds in generating the $r$-process. However, these same models and others indicate a role for production of the $p$-process nuclei via the $rp$- and $\nu p$-processes in PNS winds to generate heavy nuclei like Mo, Ru, Pd, and Te \citep{pruet2005,pruet2006,frohlich2006a,frohlich2006b,wanajo2006}. As in the case of the $r$-process, the asymptotic yields are controlled in part by the entropy of the matter and the flow dynamical expansion timescale \citep{pruet2006,fisker2009,wanajo2011a}.

One ingredient missing from these calculations is the strong magnetic fields that may accompany neutron star births.  \cite{duncan_thompson} and \cite{thompson_duncan} suggested that magnetar-strength fields $\gtrsim10^{14}$\,G might naturally arise during the collapse of rapidly-rotating massive star cores. In a study of the observed population of magnetars, \protect{\cite{woods_thompson}} argue that $\sim10$\% of all neutron stars are thought to be born with such high fields.  \cite{thompson94} argued that magnetar-strength fields $\gtrsim10^{14}$\,G would dominate PNS wind dynamics, and  \cite{qian_woosley} suggested that they might affect the $r$-process. \cite{thompson03} made a first estimate of the importance of magnetar-strength fields for nucleosynthesis, arguing that magnetic fields would trap matter close to the PNS in a helmet streamer configuration \citep{pneuman1971}, allowing the matter to reach much higher entropy before the thermal pressure gradient exceeded the magnetic tension force, precipitating dynamical ejection. \cite{thompson03} predicted a robust $r$-process in these eruptions, but the estimates made assumed wind density profiles and dynamical expansion times taken from freely expanding (non-magnetic) wind solutions, and so the conclusions were necessarily speculative. Here, we present calculations of axisymmetric MHD neutrino-heated winds from PNSs with strong surface dipole magnetic fields that self-consistently capture the neutrino heating/cooling and time-dependent dynamics. 

Although the simulations we present are the first to explore the  potential for nucleosynthesis in dynamically trapped and ejected matter near magnetized PNSs, there is already a substantial literature on the importance of the combined effects of rapid rotation and strong magnetic fields  --- ``millisecond (ms) proto-magnetars" --- under various approximations. Winds from magnetars with ms spin periods have been investigated both as a potential GRB central engine and as a potential site for the $r$-process. In particular, \cite{tcq} and  \cite{metzger07,metzger08} developed one-dimensional split-monopole models of magneto-centrifugally accelerated PNS winds in order to assess early PNS spindown as an energy source in GRBs (see \citealt{metzger11}). \cite{metzger08} showed that it is possible for ms monopole magnetar winds to achieve favorable conditions for the $r$-process via the action of strong magneto-centrifugal acceleration, which produces a low-entropy, short dynamical expansion timescale, and high-neutron fraction outflow for $P\lesssim1$\,ms.\footnote{Low electron fraction outflows are also found in the early jets from magnetorotational core collapse models \citep{Winteler2012,Nishimura2015,Nishimura2017}.}  Complementary to monopole models, numerical simulations of dynamical ms magnetar winds with surface dipole fields by \cite{bucciantini06,bucciantini08,bucciantini09} employed an adiabatic equation of state and focused on magnetar spindown and jet formation relevant to GRBs. In addition, \cite{vlasov,vlasov2017} have recently explored the importance of magneto-centrifugal acceleration and non-spherical areal divergence of flow streamlines in static force-free magnetic field configurations. Thus, the calculations done so far either assess magneto-centrifugal slinging and neutrino heating in a split-monopole geometry, and/or the effects of a static strong dipole magnetic field, or they ignore the neutrino microphysics and study relativistic MHD spindown in the adiabatic limit. They do not account for both neutrino heating/cooling and strong, but dynamical magnetic fields together in a self-consistent MHD simulation. An important exception is the work of \cite{komissarov_barkov} who did 2D simulations of neutrino-heated ms-magnetar winds. They found equatorial dynamical ejections similar to those we report below, but focused their work on jet production, dynamics, mass-loading, and time-dependence/variability relevant to the theory of magnetar-powered GRBs and their supernovae.

The purpose of this paper is to present first results from dynamical calculations for non-rotating proto-magnetar winds focused on their nucleosynthesis throughout the cooling epoch. We find that the non-relativistic magnetically-dominated magnetosphere subject to neutrino heating is unstable, with periodic plasmoid ejections from the equatorial closed-zone. As we show below, this trapped matter achieves high entropy before eruption and thermodynamic conditions that imply a successful heavy-element $r$-process if the flow is neutron-rich. In Section \ref{section:nucleosynthesis} we briefly review aspects of the $r$- and $\nu p$-processes. In Section \ref{section:numerical} we discuss our numerical model, including its limitations and our approximations. In Section \ref{section:results} we present our results, including an estimate of the total amount of material ejected above the threshold for heavy element nucleosynthesis (Section \ref{section:mass}), and a comparison with the simplified analytic model of \cite{thompson03} (Section \ref{section:analytic}). Section \ref{section:discussion} provides a discussion and conclusion.

\section{Nucleosynthesis in PNS Winds}
\label{section:nucleosynthesis}

\subsection{The r-Process}
\label{section:rprocess}

As hot material expands from the surface of the PNS it cools, first allowing $\alpha$-particles to form at around $0.5$\,MeV, and then heavier elements via the rate-limiting $\alpha$-process reaction $^4{\rm He}(\alpha n,\gamma)\,^9{\rm Be}(\alpha,n)\,^{12}{\rm C}$ and subsequent $\alpha$ captures. If the medium is neutron-rich, the $r$-process then proceeds at lower temperatures via rapid neutron captures. The nuclear flow moves along the neutron-rich side of the valley of $\beta$-stability and at sufficiently large scales the neutron-rich nuclei decay to their primary stable isobar. The asymptotic abundances are determined by the free neutron-to-seed ratio after the $\alpha$ process \citep{woosley_hoffman,meyer94,witti94,hoffman97}.

That non-rotating, non-magnetic PNS winds fail to achieve conditions necessary for production of the 3rd $r$-process peak is quantified in terms of a single figure of merit derived from the $^4{\rm He}(\alpha n,\gamma)\,^9{\rm Be}(\alpha,n)\,^{12}{\rm C}$ reaction \citep{hoffman97}:
\beq
\zeta_{\rm crit}=\frac{S^3}{Y_e^3 \,t_{\rm dyn}}\simeq8\times10^9\,\,({\rm k_B\,\,baryon^{-1}})^3\,\,{\rm s^{-1}},
\label{zetacrit}
\eeq
where $S$ is the entropy per baryon,\footnote{We use units of $k_{\rm B}$ baryon$^{-1}$ for $S$ throughout.}  $Y_e$ is the electron fraction, and $t_{\rm dyn}$ is the dynamical expansion timescale. For $\zeta>\zeta_{\rm crit}$, the nuclear flow can proceed to the actinides, but for $\zeta<\zeta_{\rm crit}$, the nucleosynthesis halts at smaller mass numbers (see also \citealt{meyer_brown}). We note that the numerical value of $\zeta_{\rm crit}$ varies in the calculations of \cite{hoffman97}, depending on the underlying equation of state, and with a more complicated dependence on $Y_e$ near $\simeq0.5$. Nevertheless, $\zeta_{\rm crit}$ serves as a useful guide in diagnosing the simulations presented in Section \ref{section:results}.

In non-rotating, non-magnetic PNS winds, $Y_e$ is set by the relative luminosities and energies of the electron-type neutrinos, $L_{\nu_e}/L_{\bar{\nu}_e}$ and $\langle\varepsilon_{\nu_e}\rangle/\langle\varepsilon_{\bar{\nu}_e}\rangle$, through the charged-current processes $e^-+p\leftrightarrow n+\nu_e$ and $e^++n\leftrightarrow p+\bar{\nu}_e$ \citep{qian_woosley}. Models of cooling PNSs imply that the medium may be moderately neutron-rich with $Y_e\simeq0.4-0.5$, but possibly proton-rich at both early and later times during the cooling epoch  \citep{qian_woosley,fischer2010,hudepohl2010,roberts12_cooling,roberts12_wind, roberts_PNS}. Non-rotating, non-magnetic wind models show that $S$ ranges from $\sim50$ to at most $200$ while  $t_{\rm dyn}$ ranges from $0.01-0.1$\,s, implying that 
 \beq
 \zeta\simeq2\times10^8\left(\frac{S}{100}\right)^3\left(\frac{0.01\,{\rm s}}{t_{\rm dyn}}\right)\left(\frac{0.5}{Y_e}\right)^3, 
 \label{zeta_typical}
 \eeq
significantly less than $\zeta_{\rm crit}$. As the PNS cools, both $S$ and $t_{\rm dyn}$ increase, but the wind does not evolve into a state with $\zeta>\zeta_{\rm crit}$ \citep{qian_woosley,otsuki,tbm}. The implication is that either PNSs produce only the 1st $r$-process abundance peak and another mechanism is responsible for heavy $r$-process production (e.g., neutron star binary mergers; e.g., \citealt{eichler1989,freiburghaus1999,korobkin2012}), or that additional physics allows some or all PNS births to produce the heavier nuclei. Models including General Relativity \citep{cardall_fuller}, extra heating sources \citep{qian_woosley,suzuki_nagataki,metzger07}, magneto-centrifugal acceleration \citep{metzger08}, and non-spherical areal divergence in strong static magnetic field configurations \citep{vlasov,vlasov2017} have all been explored in an effort to bridge the gap between typical values of $\zeta$ as in equation (\ref{zeta_typical}) and $\zeta_{\rm crit}$ in equation (\ref{zetacrit}).

\subsection{The p-Nuclei}
\label{section:pnuclei}

The production of the $p$-nuclei in proton-rich PNS winds has been studied by a number of authors, focusing on either the very early time ejecta or the role of neutrinos \citep{hoffman1996,pruet2005,pruet2006,frohlich2006a,frohlich2006b,wanajo2006,wanajo2011a}. There have not been systematic explorations of the complete wind parameter space analogous to the $r$-process surveys of \cite{meyer_brown} or \cite{hoffman97}, which cover a broad range of entropy, dynamical timescale, and electron fraction. However, \cite{pruet2006} report results for winds with entropy significantly larger than their fiducial wind model, which has $S\simeq55-77$, $Y_e\simeq0.54-0.56$, and synthesizes elements up to $^{102}$Pd. For 2 and 3 times higher entropy, they find that $\nu p$-process nucleosynthesis extends up to $^{120}$Te and $^{168}$Yb, respectively. These results imply that the entropy of the medium has a direct impact on the asymptotic yield. 

\begin{figure*}
\centerline{
\includegraphics[width=5.5cm]{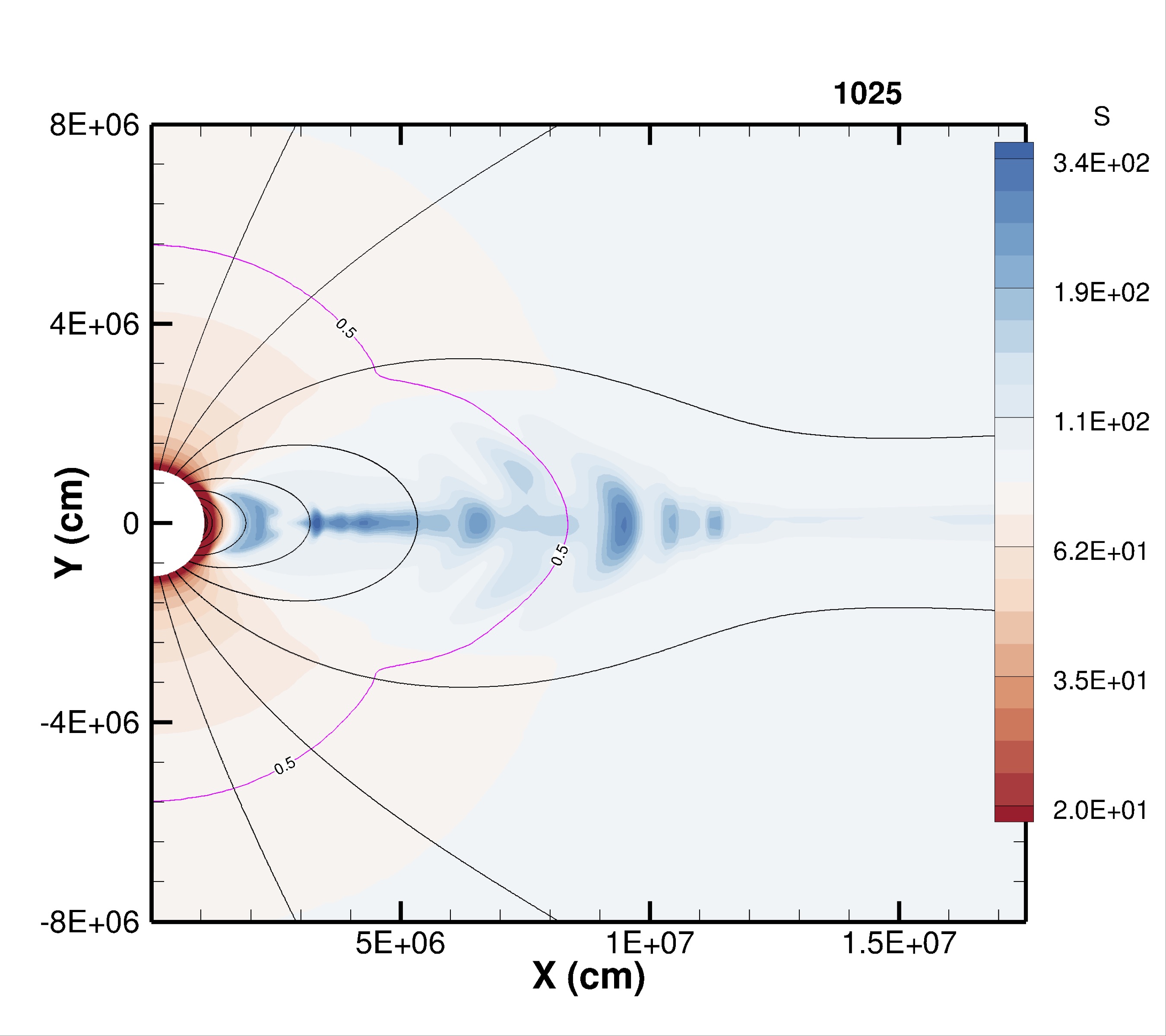}
\includegraphics[width=5.5cm]{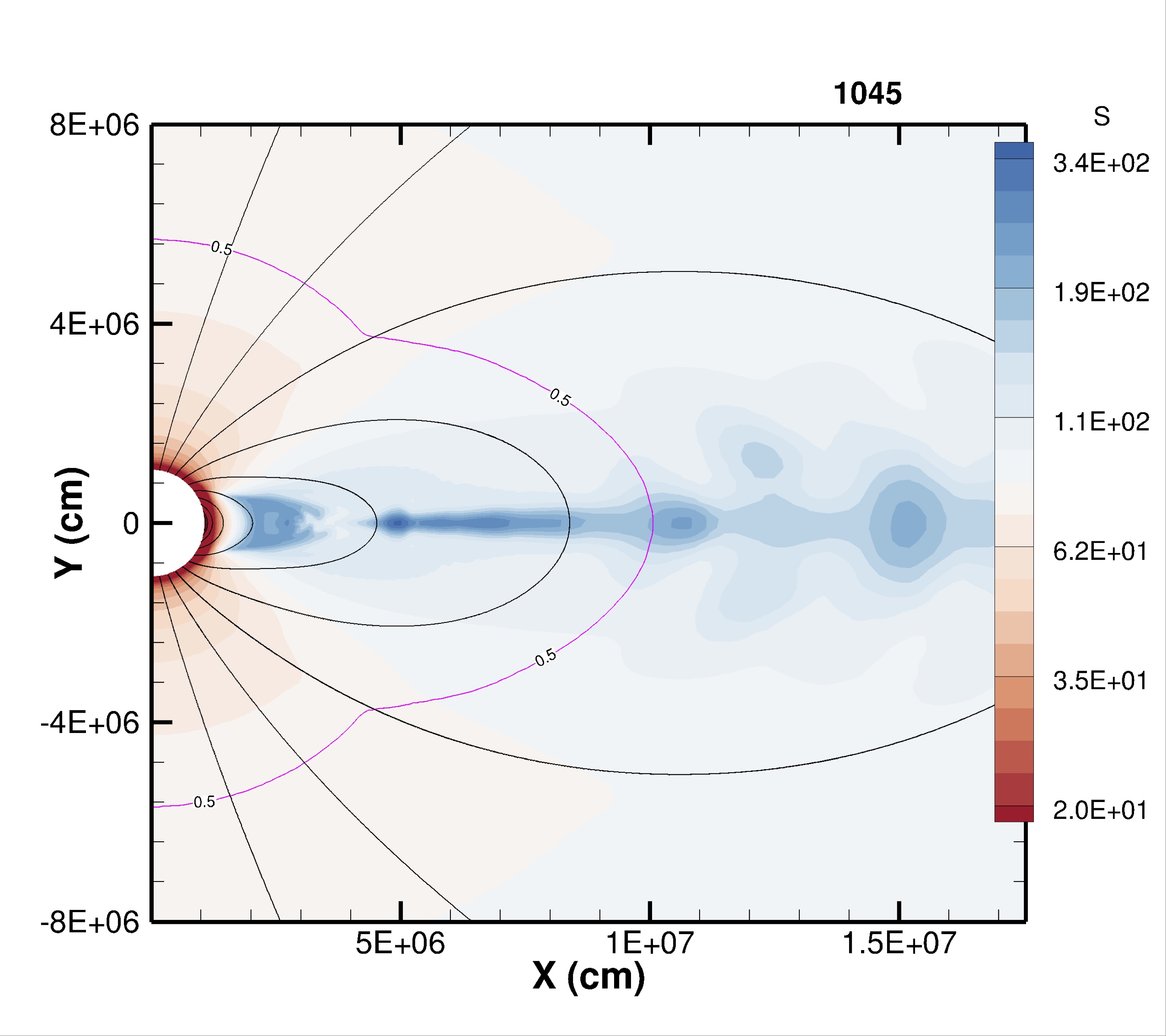}
\includegraphics[width=5.5cm]{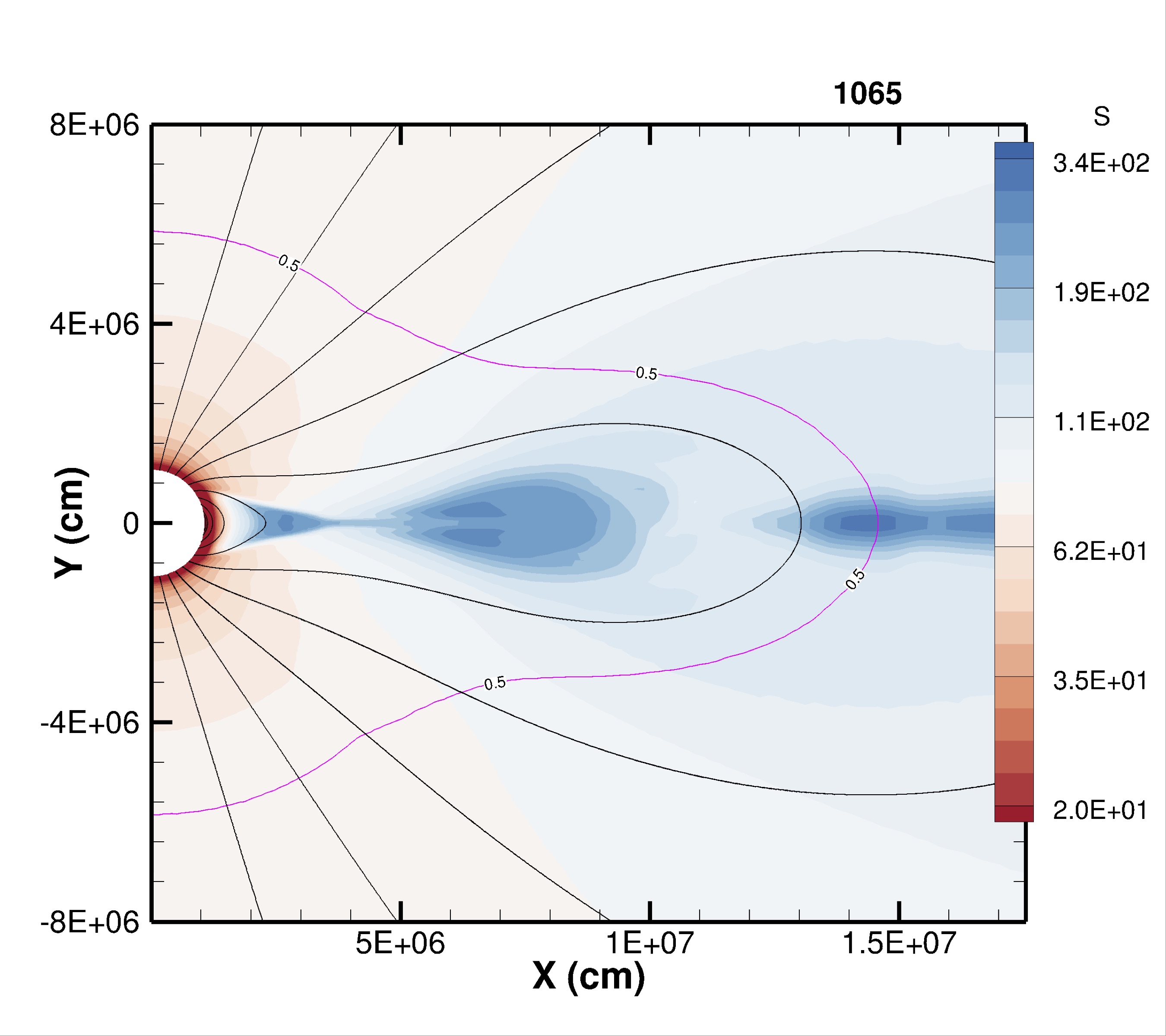}}
\centerline{
\includegraphics[width=5.5cm]{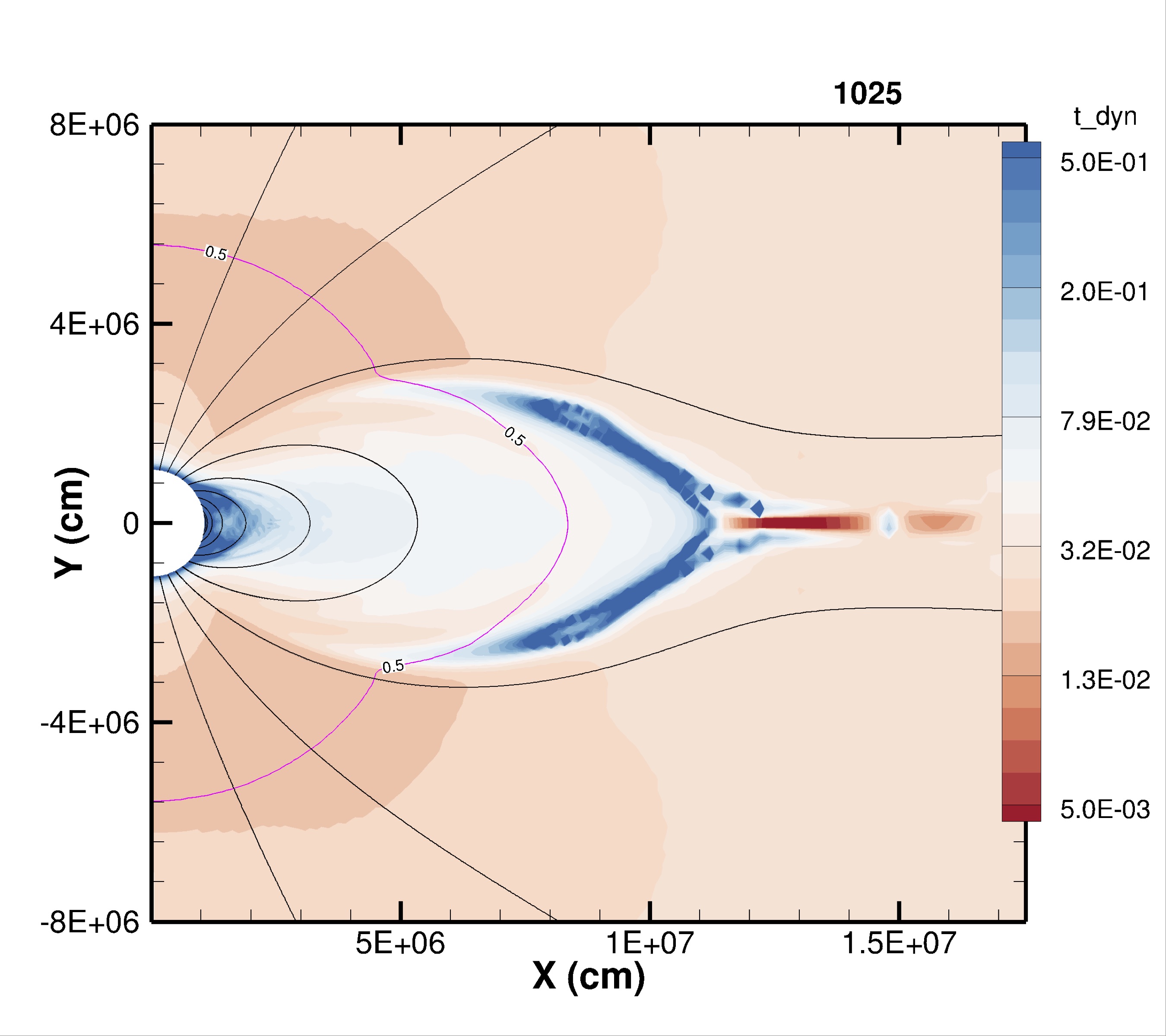}
\includegraphics[width=5.5cm]{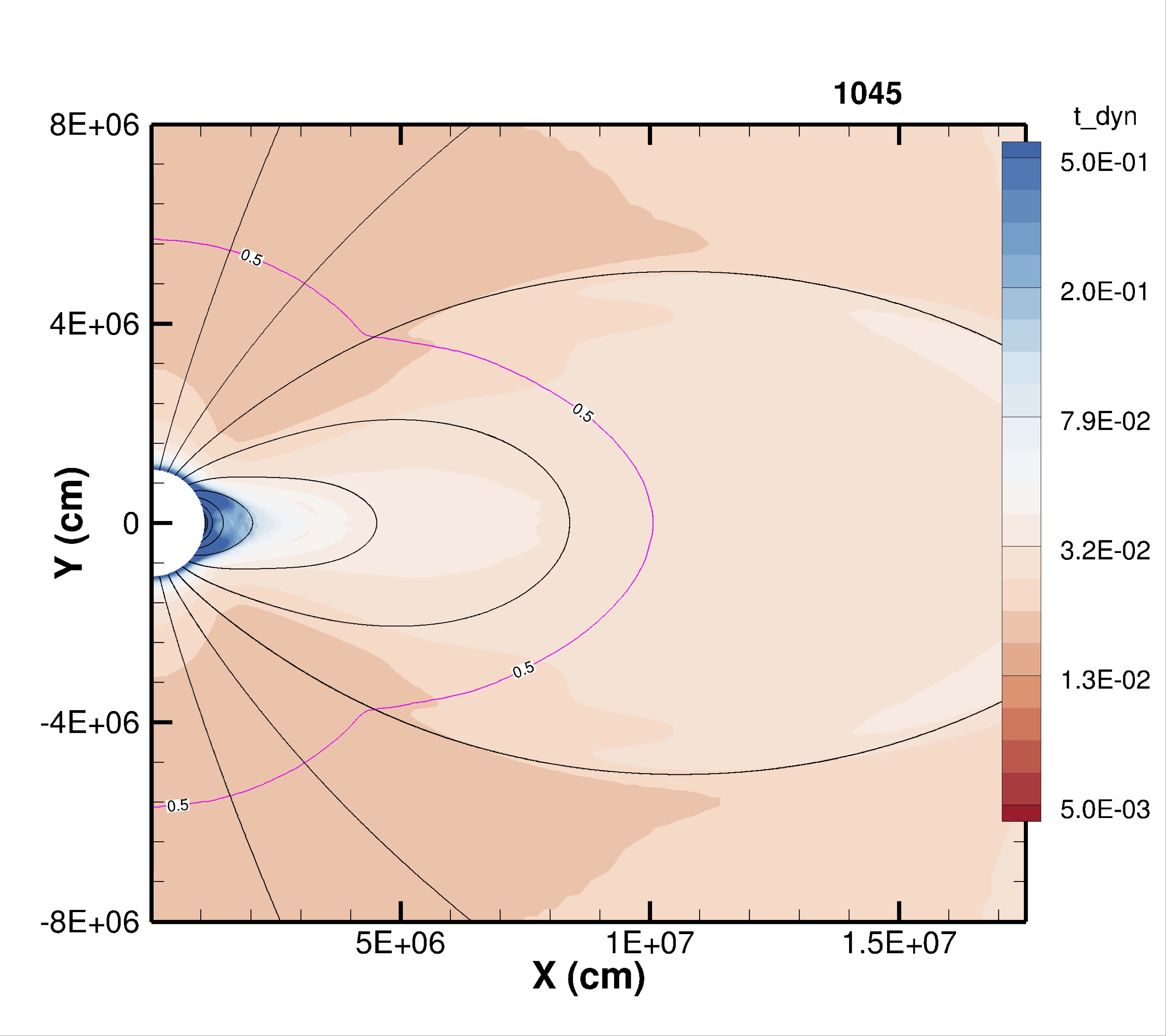}
\includegraphics[width=5.5cm]{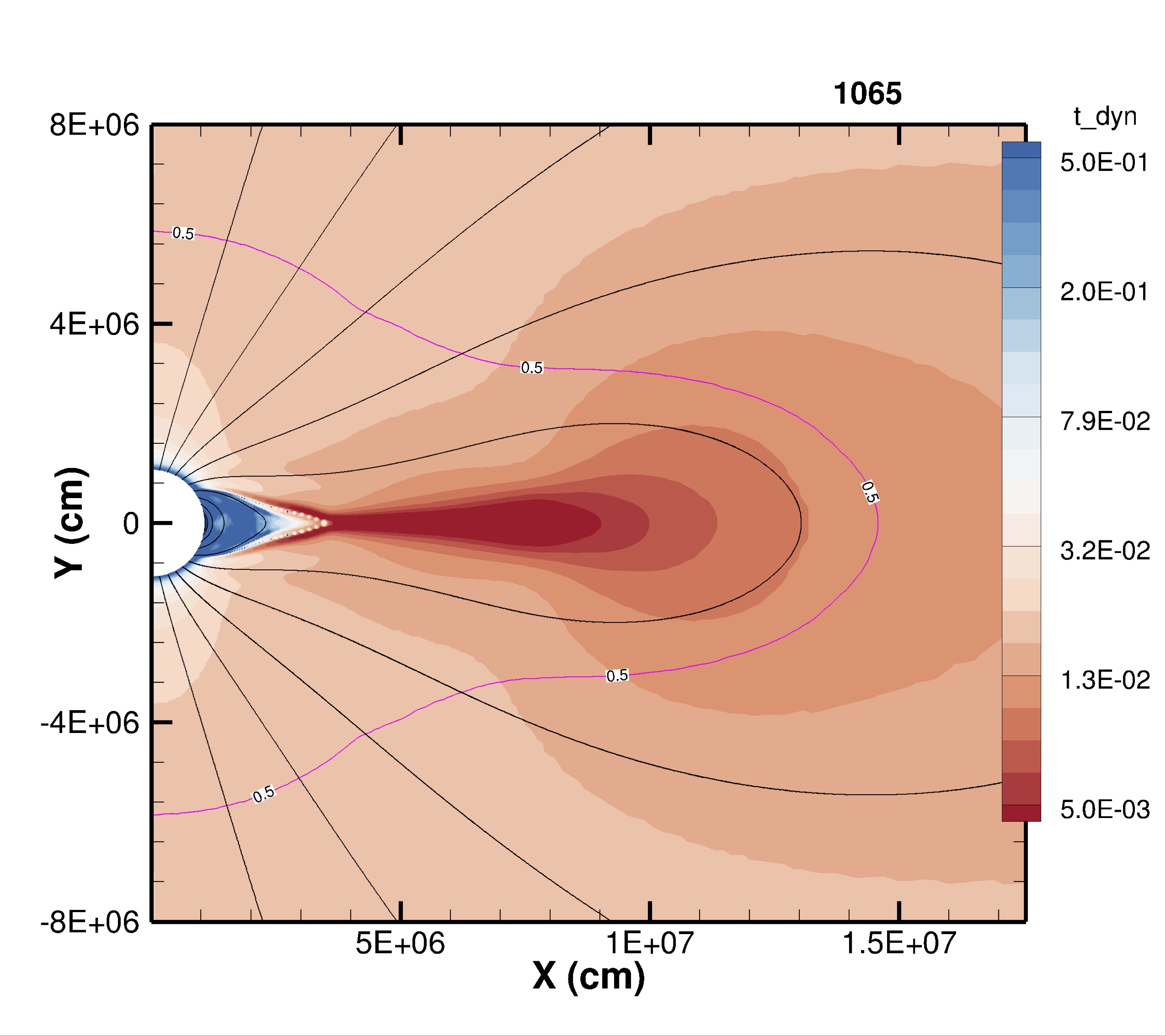}}
\centerline{
\includegraphics[width=5.5cm]{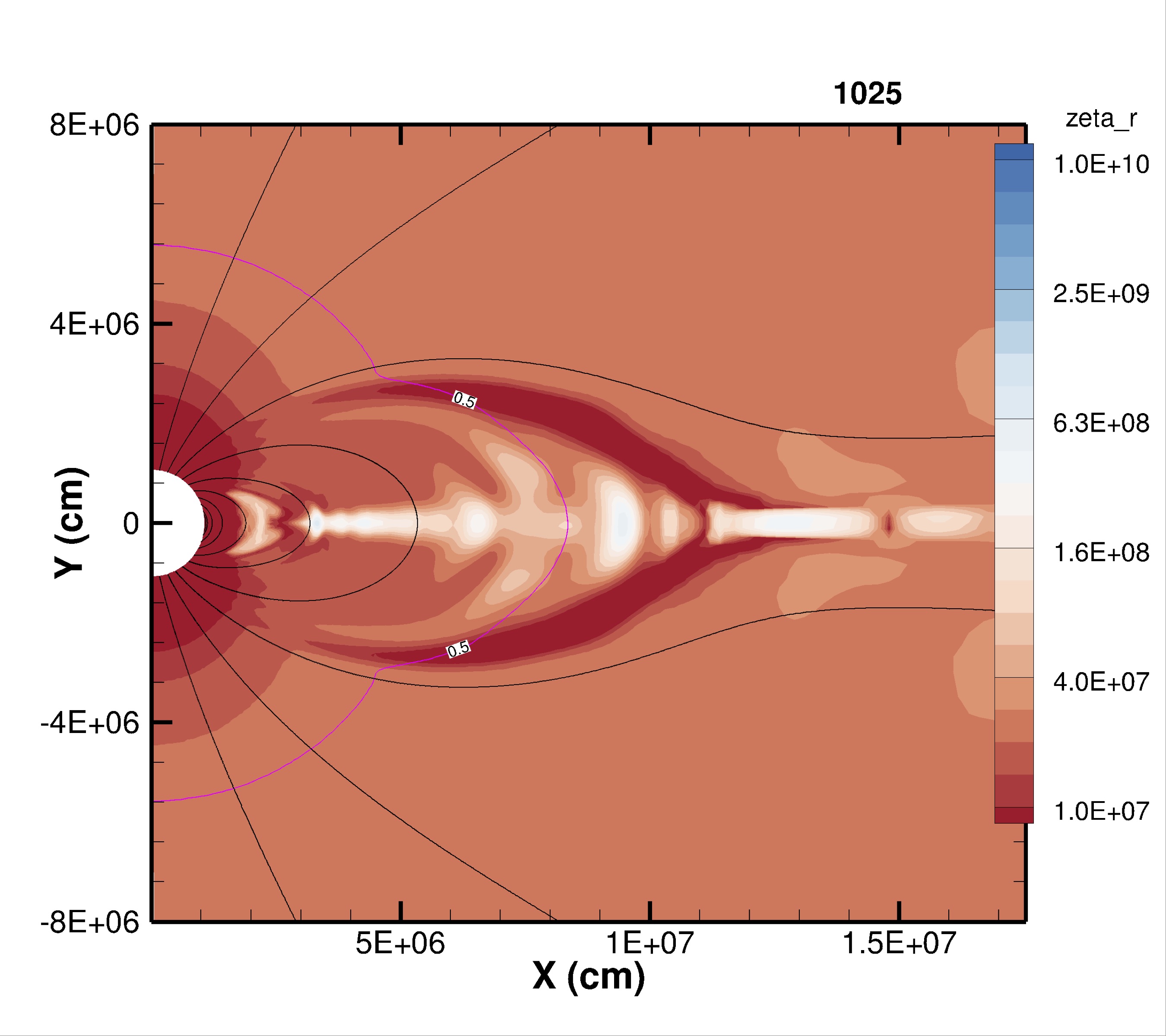}
\includegraphics[width=5.5cm]{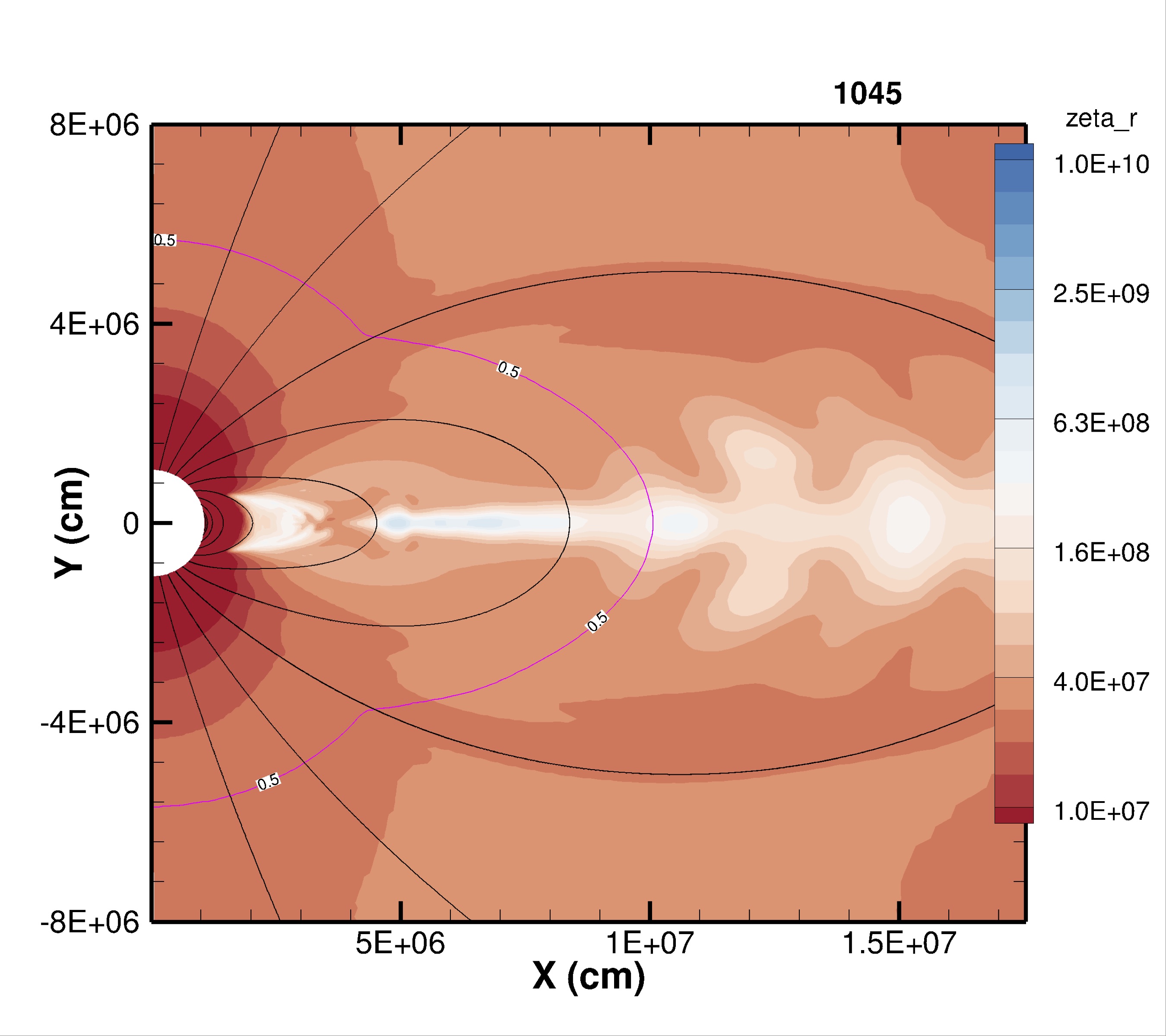}
\includegraphics[width=5.5cm]{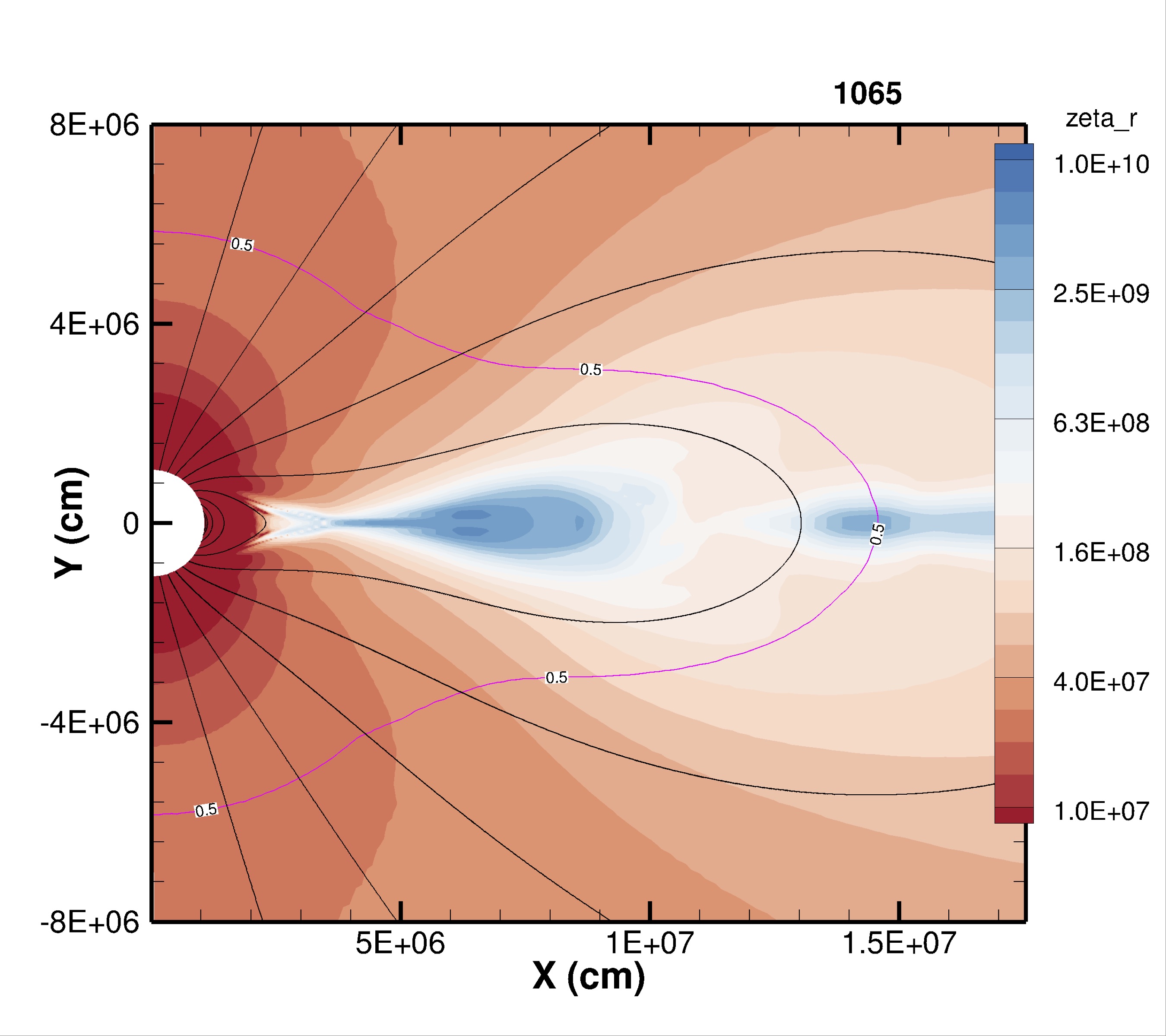}}
\caption{Snapshots of the wind entropy $S$ (top panels), dynamical timescale $t_{\rm dyn}=r/V_r$ (middle panels), and $\zeta_r$ (eq.~\ref{zetar}; bottom panels) at times 1025, 1045, and 1065\,ms in a simulation with $L_\nu =8\times10^{51}$\,ergs/s and $B=10^{16}$\,G, showing the emergence of a high-entropy plasmoid. The black lines indicate the magnetic field. The magenta line denotes the $T=0.5$\,MeV surface. This particular ejection episode was chosen as a representative example. The mass ejected as a function of $\zeta$ and time for this model is shown in Figure \ref{figure:mass} (left panel). Histograms of mass as a function of $\zeta$ for a specific ejection is shown in Figure \ref{figure:hz}.}
\label{figure:highl}
\end{figure*}

\section{Numerical Model}
\label{section:numerical}

We use the publicly available non-relativistic ideal MHD code ZEUS-MP \citep{stone_norman1,stone_norman2} with a staggered mesh grid consisting of $400$ logarithmically spaced radial zones from the PNS surface at $R_{\rm NS}=10.7$\,km out to $5\times10^3$\,km and $200$ evenly spaced zones in $\theta$. We initialize all models with a 1D spherically symmetric wind model and a dipole field with strength ranging from $B=0$ to $10^{16}$\,G. All models have a PNS mass of $1.4$\,M$_\odot$\footnote{Higher mass PNSs (e.g., up to $\sim2$\,M$_\odot$) produce higher entropy and lower mass loss rate, all other parameters held constant \citep{qian_woosley,tbm}).} and are non-rotating. The latter is particularly important since even a modest PNS spin period of $P\sim3-5$\,ms can affect the dynamical expansion timescale, which directly affects $\zeta$ \citep{metzger08,vlasov,vlasov2017}. We return to the issue of rotation in Section \ref{section:discussion}.

We find that the behavior is dependent on resolution in both radial and $\theta$ directions. Insufficient resolution in either direction can lead to qualitatively different behavior. In particular, poor resolution in the radial direction may lead to erroneous mass loss rate and entropy, whereas under-resolving the co-latitudinal direction may inhibit the numerical reconnection events that are essential to the dynamics of the episodic ejections seen in this work, leading to instead a relatively steady solution. Thus, care must be taken. We conducted a resolution study before settling on our current grid parameters. In particular, we found that decreasing the angular resolution by a factor of 2 from our fiducial resolution led to {\it steady} non-episodic flow, without the large increase in entropy reported below. We have tested higher resolution and found that our reported results do not change qualitatively.

All calculations are initialized with a fixed PNS neutrino luminosity. We index our models with the anti-electron neutrino luminosity $L_{\bar{\nu}_e}$, varying it from $10-0.1\times10^{51}$ \,ergs/s. For a given $L_{\bar{\nu}_e}$, we assume that the electron-type neutrino luminosity is $L_{\nu_e}=L_{\bar{\nu}_e}/1.3$ (e.g., \citealt{tbp}). The range in neutrino luminosity considered is meant to be representative of different times in the PNS cooling epoch ranging from $\sim1-100$\,s after PNS formation (e.g., compare with \citealt{pons1999}). 

Rather than attempting to model the time-evolution of $\lumanue$, we calculate individual $\sim1$\,s snapshots for a given surface magnetic dipole field $B$ and fixed luminosity. Our primary reason for doing this is that we are using non-relativistic MHD. In tests, we find that for a given fixed $B$, as the neutrino luminosity decreases, the net mass outflow rate decreases and the system rapidly becomes more relativistic, eventually having magnetosonic velocities in excess of $c$, as expected from analytic arguments  and in keeping with our intuition that the flow should approach the ``pulsar"-like Poynting flux dominated phase as the PNS cools \citep{tcq,metzger07,metzger11}. Because our results are not valid in the relativistic regime, we focus on snapshots for given $B$ and $\lumanue$ combinations such that the magnetosonic speeds are subluminal. We return to how these snapshots fit into a global view of PNS winds in Section \ref{section:discussion}. 

The neutrino heating rates depend on both neutrino luminosity and average neutrino energy. We opt to hold the latter constant at $\langle\varepsilon_{\nu_e}\rangle=11$ and $\langle\varepsilon_{\bar{\nu}_e}\rangle=13$\,MeV for all $\lumanue$. The primary importance of the neutrino energies is to set the total heating rate and $Y_e$ of the outflow. Since we vary the neutrino luminosity widely, and since we only model snapshots throughout the cooling epoch, and since we do not calculate $Y_e$ self-consistently, this approximation is acceptable. Future calculations will explore the time-dependence of $\lumanue$, $\lumnue$, $\langle\varepsilon_{\nu_e}\rangle$, and $\langle\varepsilon_{\bar{\nu}_e}\rangle$ from PNS cooling calculations in an attempt to model the entire wind phase self-consistently.

We make a number of simplifying assumptions to the physics. We neglect the importance of General Relativity \citep{cardall_fuller,otsuki,pruet2001,tbm}. The deeper potential well in GR combined with the gravitational redshifts entering the neutrino heating rates lead to overall higher entropy and lower mass outflow rate than the Newtonian calculations presented here. We use a simple equation of state that includes non-relativistic nucleons as an ideal gas, relativistic electron/positron pairs, and photons. The neutrino heating and cooling terms include only the charged-current reaction rates using the approximations of \cite{qian_woosley}. We thus ignore the sub-dominant contributions to the heating/cooling from inelastic neutrino-nucleon and -electron scattering and electron/positron annihilation to $\nu \bar{\nu}$ \citep{tbm}. Furthermore, we neglect the importance of strong magnetic fields on the electron/positron phase space distribution imposed by Landau quantization, which affects both the equation of state and the neutrino interactions \citep{lai1998,arras1999,duan2004,duan2005}. In addition, rather than solving a separate evolution equation for the electron fraction $Y_e$, we make the approximation that $Y_e=0.45$ is constant throughout the flow. Although $Y_e$ is critical to the nucleosynthesis of the outflow, its radial evolution is not of dominant importance to the wind dynamics; typically, $Y_e$ rapidly increases from $\sim0.1$ near the PNS neutrinosphere to its asymptotic value set by the $\nu_e$ and $\bar{\nu}_e$ energies and luminosities at $Y_e\sim0.4-0.5$ over just a few$\,-$10\,km from the PNS \citep{qian_woosley}. In addition, the formation of $\alpha$ particles is also ignored. This has the effect of somewhat increasing the radial range over which neutrino heating acts on the wind \citep{tbm}.  Finally, any potential effects of neutrino oscillations and non-standard neutrino physics (e.g., a sterile species) are not considered (e.g., \citealt{fetter2003,duan2006}). Future efforts should investigate the importance of relaxing these assumptions to the physics in dynamical MHD models of PNS winds.

Using these simplifications in the 2D calculations, we ran a number of $B=0$ cases to compare with steady-state models from the non-relativistic PNS wind calculations of \cite{tbm} using the same microphysics, and found good agreement in terms of mass outflow rate, asymptotic entropy, dynamical timescale, and velocity. The basic qualitative behaviors of earlier spherical calculations are reproduced, with quantitative differences owing to the approximations to the physics. Below, we focus on the relative comparison between these calculations and those including a dynamically important PNS magnetic field.

\begin{figure*}
\centerline{ 
\includegraphics[width=5.5cm]{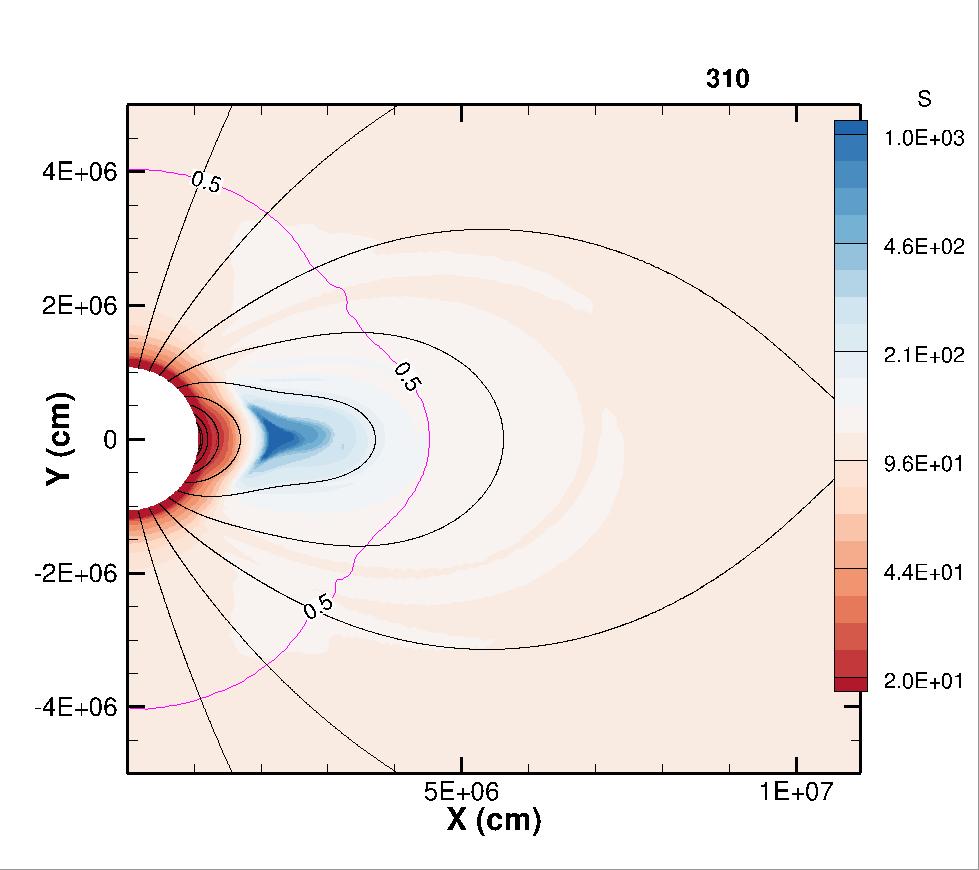}
\includegraphics[width=5.5cm]{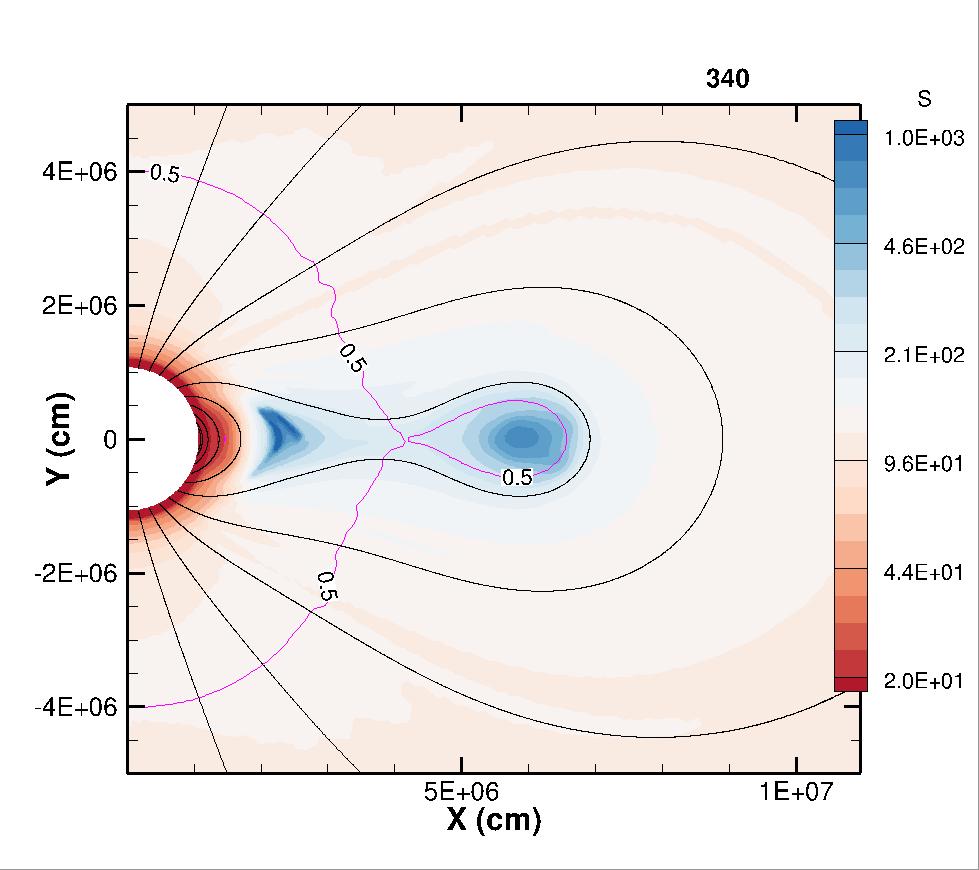}
\includegraphics[width=5.5cm]{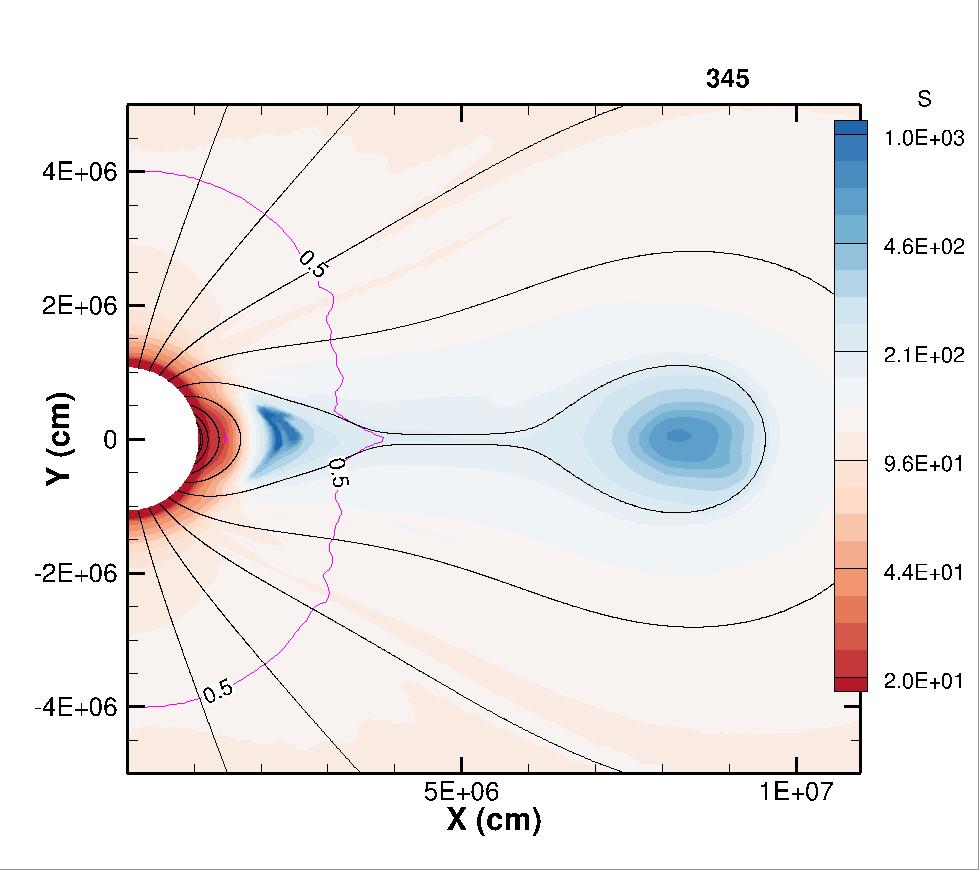}}
\centerline{ 
\includegraphics[width=5.5cm]{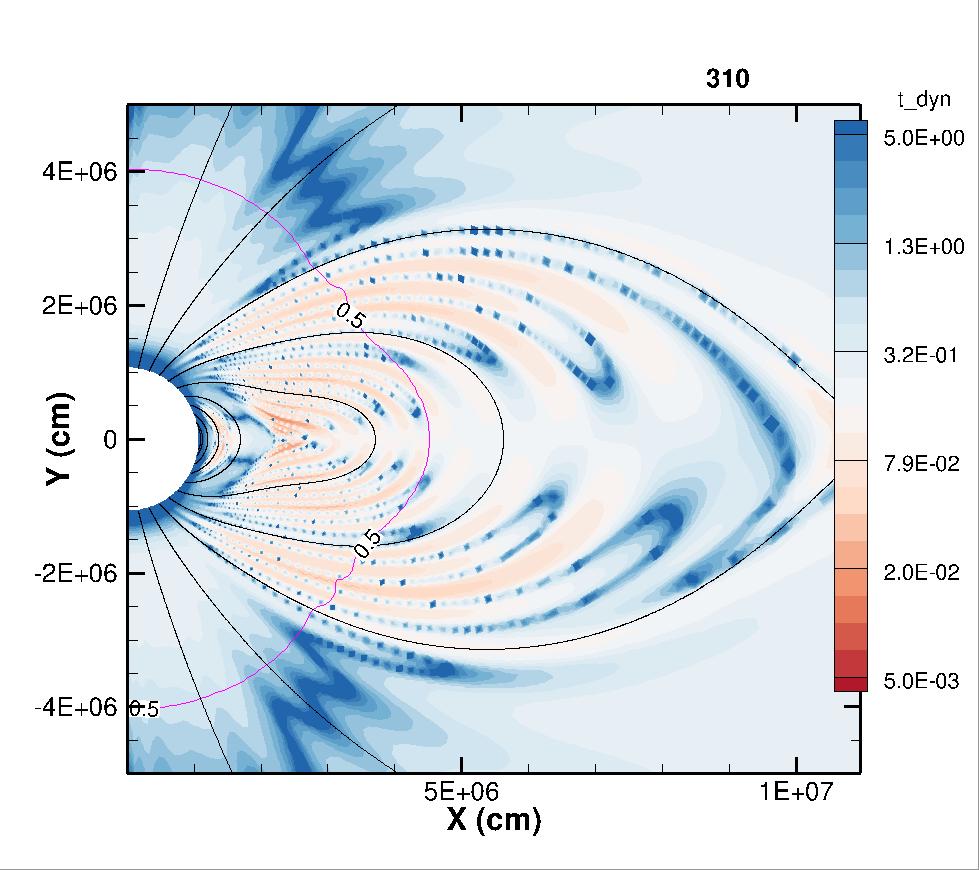}
\includegraphics[width=5.5cm]{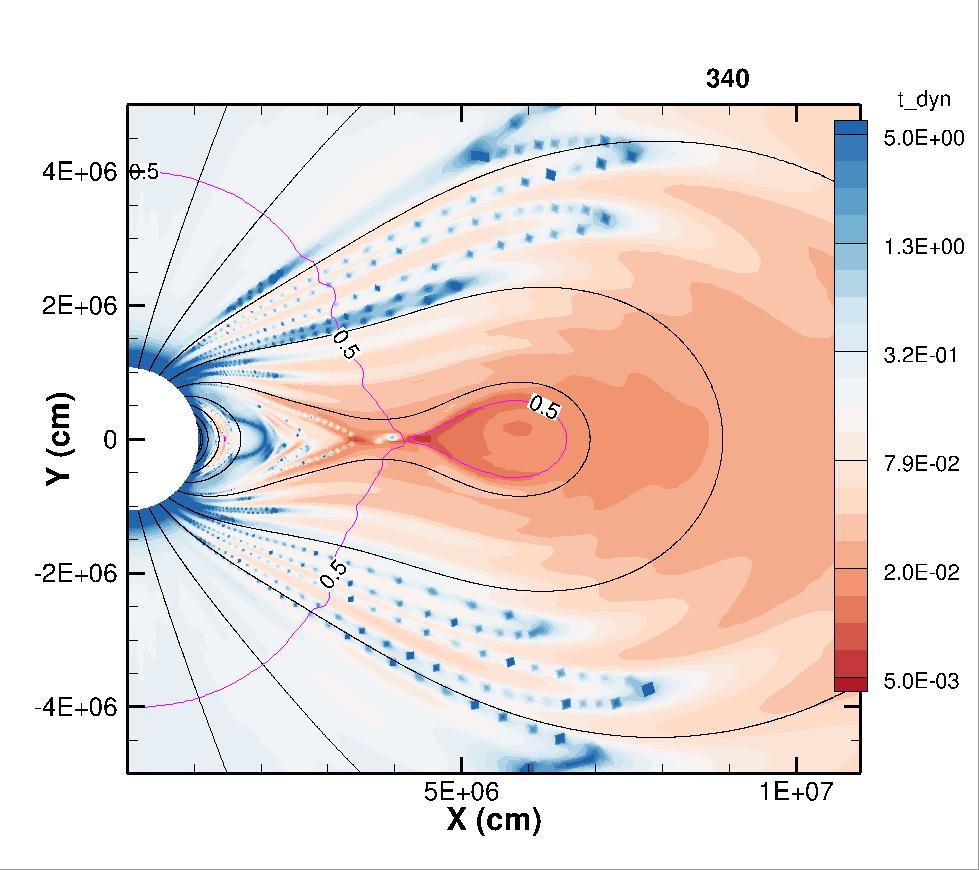}
\includegraphics[width=5.5cm]{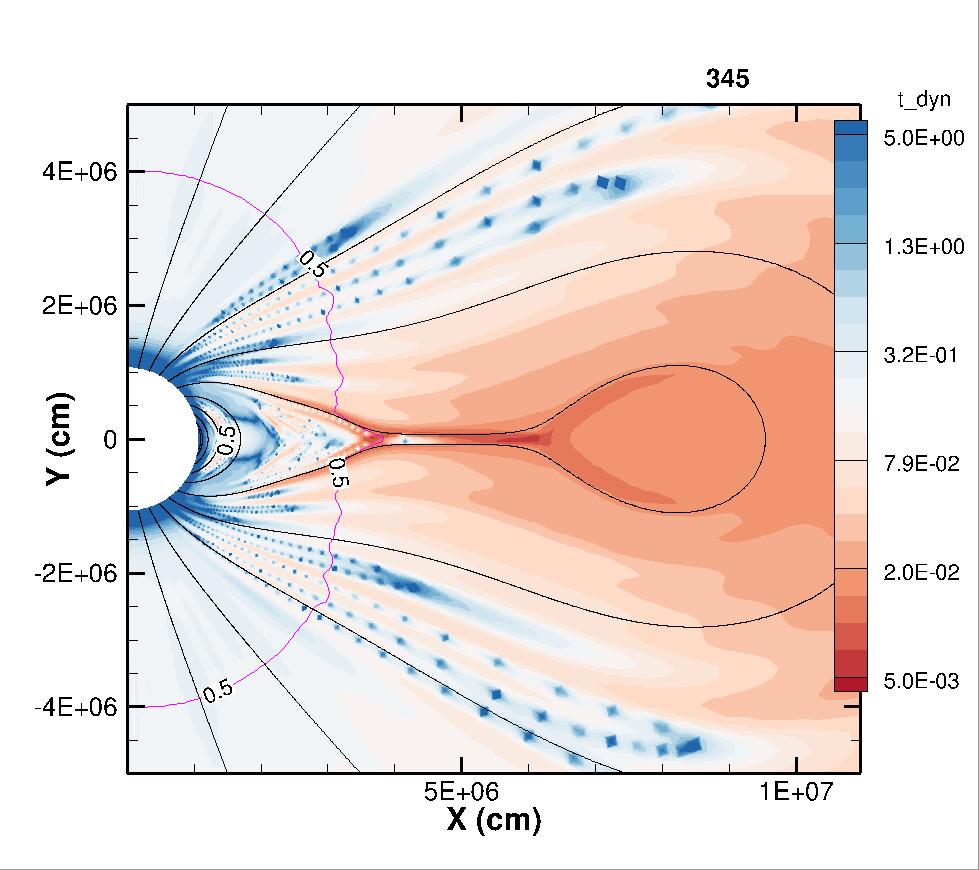}}
\centerline{ 
\includegraphics[width=5.5cm]{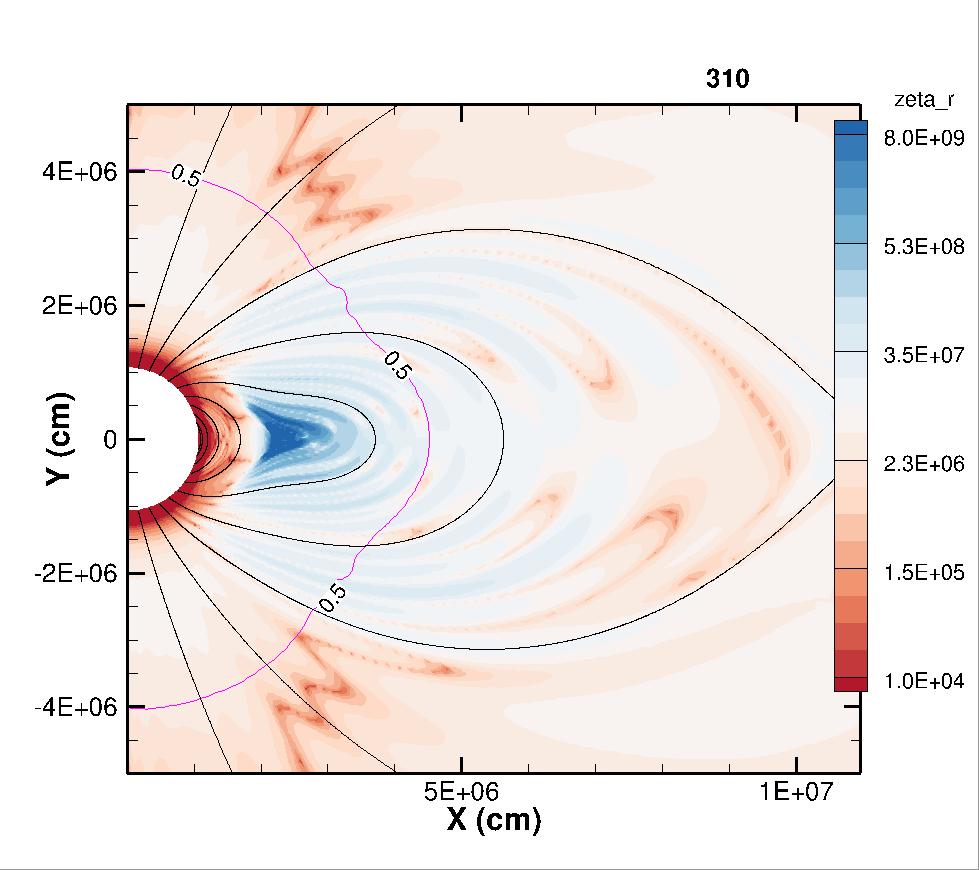}
\includegraphics[width=5.5cm]{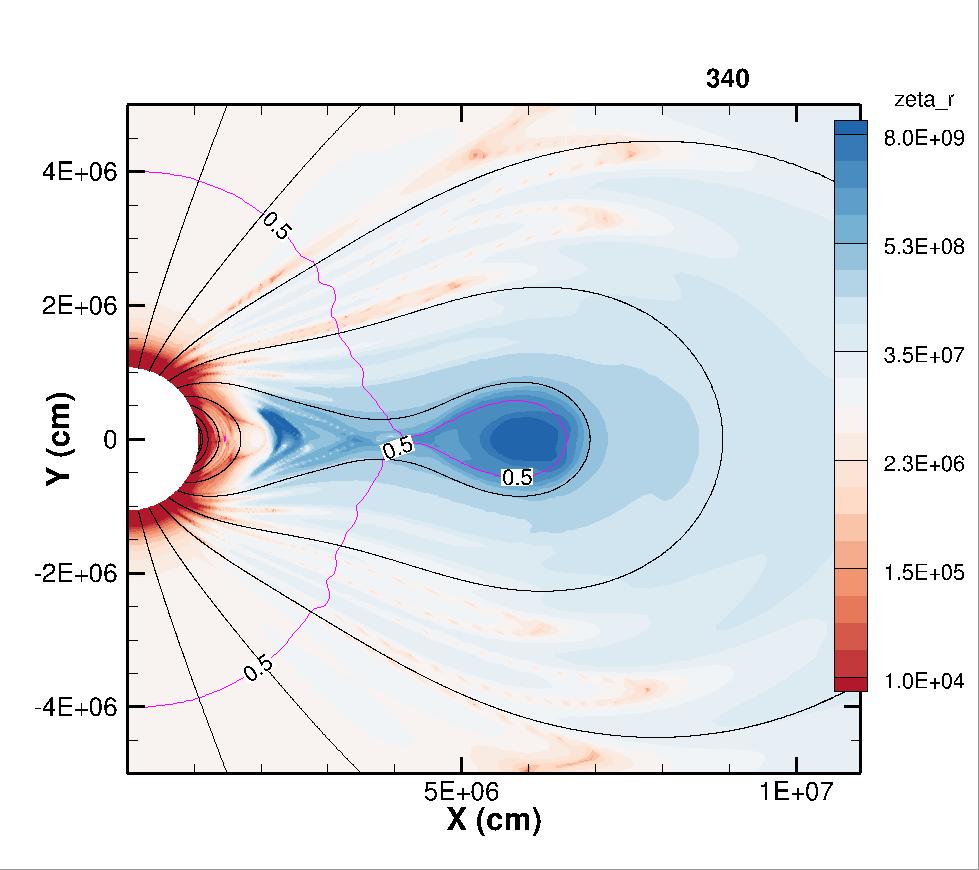}
\includegraphics[width=5.5cm]{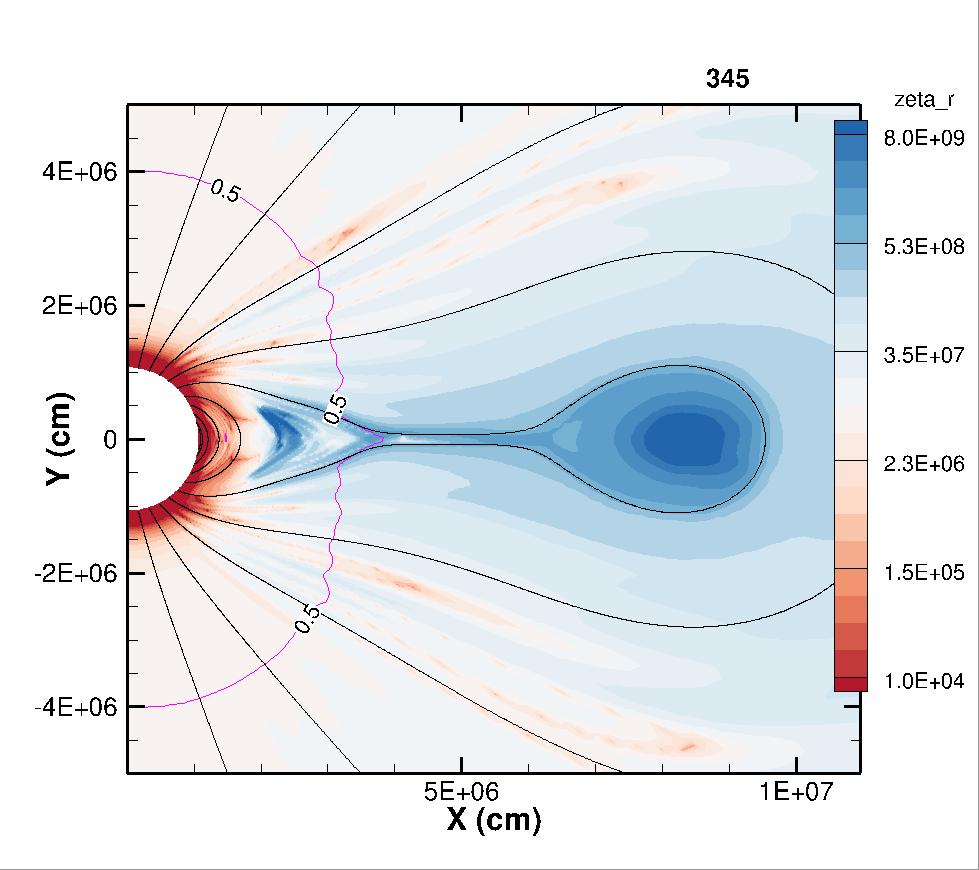}}
\caption{Snapshots of the wind entropy (top panels) and $\zeta_r$ (bottom panels) at times of 300, 345, and 350\,ms in a low neutrino luminosity simulation with $L_\nu=4\times10^{50}$\,ergs/s and $B=10^{15}$\,G, showing the emergence of a high-entropy plasmoid. Mass ejected as a function of time and $\zeta$ is shown in Figure \ref{figure:mass} (right panel).}
\label{figure:lowl}
\end{figure*}

\section{Results}
\label{section:results}

Figure \ref{figure:highl} shows results from a simulation with $B=10^{16}$\,G and high $\lumanue=8\times10^{51}$\,ergs/s, representing typical neutrino luminosities in the first $\sim1$ second after explosion (see, e.g., Figure 7 of \citealt{sukhbold2016}). The top, middle, and bottom panels of Figure \ref{figure:highl} show $S$, $t_{\rm dyn}=r/V_r$, and 
\beq
\zeta_r\equiv S^3/[Y_e^3 (r/V_r)],
\label{zetar}
\eeq
respectively, at times of $t=1025$, 1045, and 1065\,ms after the start of the simulation. The black lines indicate the orientation of the magnetic field lines. The magenta line denotes the $T=0.5$\,MeV surface. In calculating $\zeta_r$ we have assumed $Y_e=0.45$.  

As anticipated by \cite{thompson03} (see Section \ref{section:analytic}), a region of trapped material forms at the magnetic equator, in a helmet streamer configuration, within $\sim10$\,km of the PNS. An extended component to the closed zone extends over $\sim5-10$ PNS radii. The outer part of the closed zone inflates a high-entropy plasmoid that then erupts from the magnetosphere, and the cycle repeats every $\sim90$\,ms. These panels show the 10th such ejection in this simulation. After each ejection, the extended closed magnetosphere reforms rapidly via reconnection. At the equator, a high supersonic ``jet" is produced after each ejection, as formerly open streamlines are suddenly re-confined and closed, forming a converging flow at the equator and an associated entropy enhancement. Although the dynamical timescales are very short during an ejection (middle right panel), $t_{\rm dyn}$ can become as long as $\sim0.5$\,s in the closed zone after an ejection as the magnetosphere reforms and $V_r$ decreases. The entropy is significantly enhanced in the closed zone and a portion of the high-entropy material escapes in each ejection.  Whereas the spherical $B=0$ case yields an asymptotic entropy of $S\simeq70$, with magnetic fields the entropy at the magnetic equator reaches from $S\sim350$ up to a maximum of $\sim470$. As the ejection occurs, the dynamical timescale reaches $t_{\rm dyn}\simeq5\times10^{-3}$\,s, so that $\zeta_r\simeq10^{10}$, above the threshold $\zeta_{\rm crit}$ in equation (\ref{zetacrit}).

Figure \ref{figure:lowl} shows a much lower $\lumanue=4\times10^{50}$\,ergs/s and $B=10^{15}$\,G example of the dynamical ejection of a high-entropy plasmoid. This $\lumanue$ corresponds to a time $\sim10$\,s after PNS formation. The panels are arranged similarly to Figure \ref{figure:highl} and the time snapshots occur at 310, 340, and 345\,ms. Again, the panel sequence from left to right shows the emergence of a single plasmoid. The entropy in the emerging equatorial plasmoid is dramatically increased over the background. The high latitude regions of the wind reach an asymptotic entropy of only $\simeq150$, whereas the material in the plasmoid has reaches $S\simeq1200$. The $B=0$ calculation yields an entropy of $S=116$. The bottom panels show $\zeta_r$ (compare with eq.\ \ref{zetacrit}) The high entropy toroidal plasmoid exceeds the threshold for production of the 3rd peak $r$-process nuclei if the medium is neutron rich, and a large region around the equator has enhanced $\zeta$ relative to the high-latitude wind. 

\begin{figure*}
\centerline{\includegraphics[width=8.75cm]{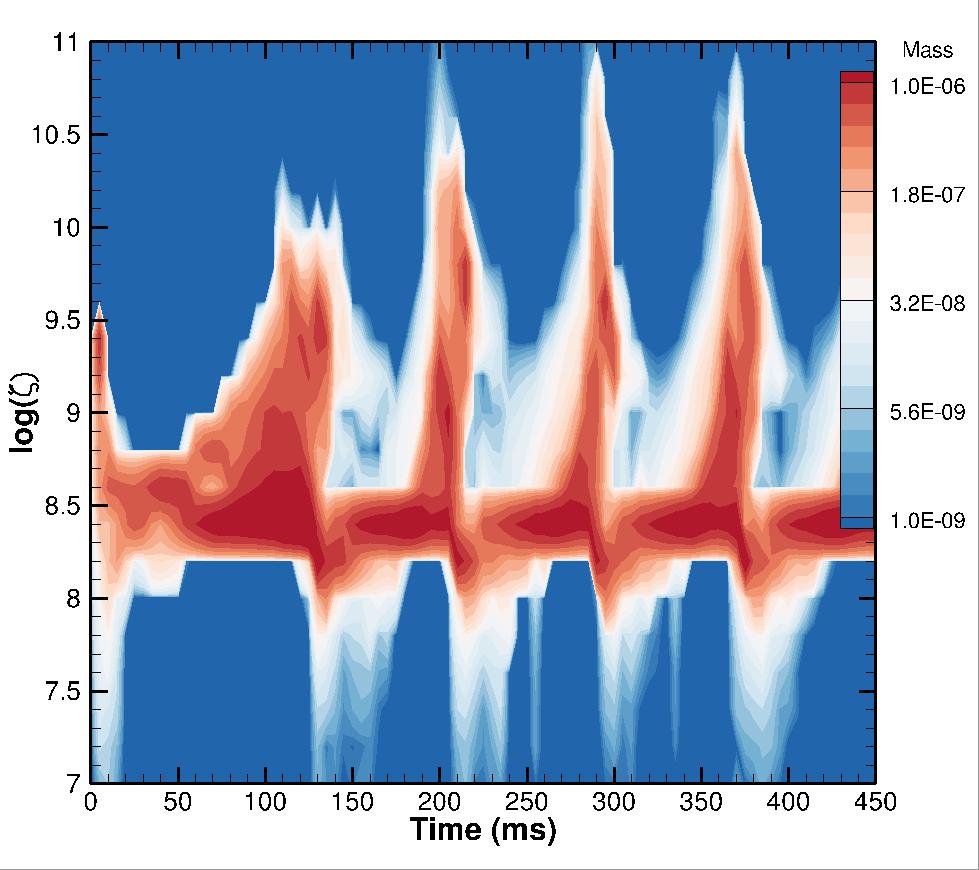}\includegraphics[width=8.75cm]{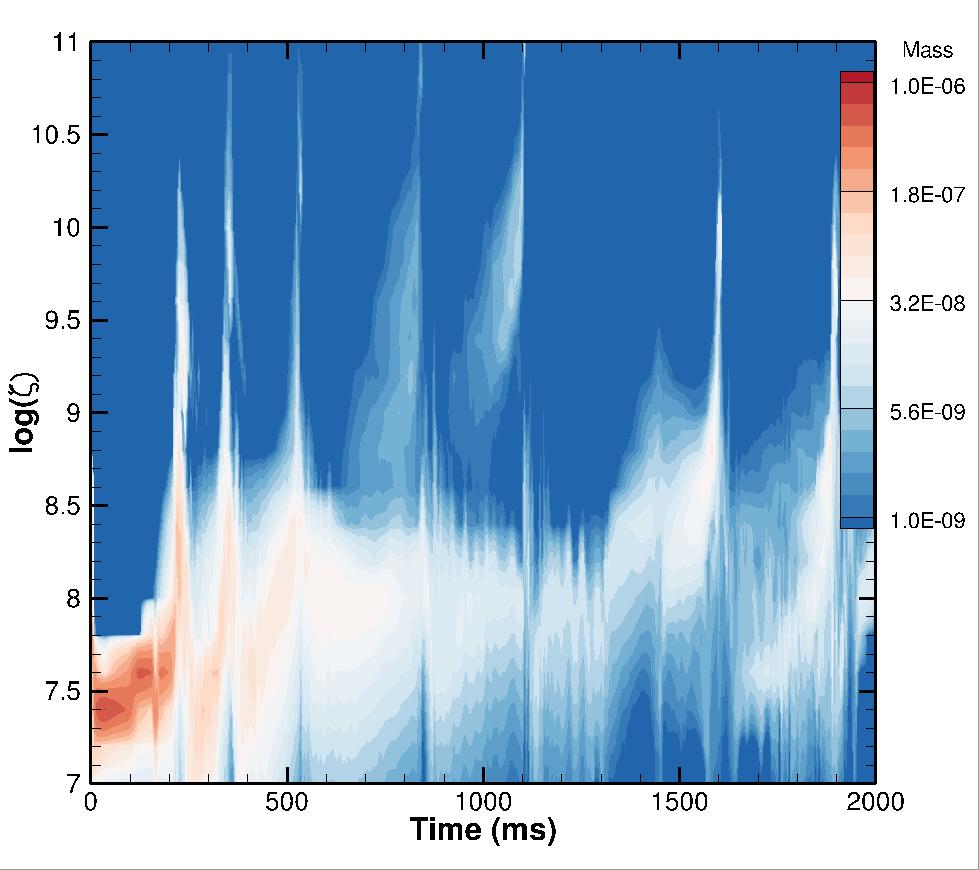}}
\caption{$\zeta$ as a function of time for the mass in the computational domain between $0.1\leq T\leq0.5$\,MeV for models with $\lumanue=8\times10^{51}$\,ergs/s (left; see Fig.~\ref{figure:highl}) and $4\times10^{50}$\,ergs/s (right; see Fig.~\ref{figure:lowl}). In the high luminosity model we observe periodic ejections on timescale $\simeq90$\,ms for the entire $\sim1$\,s duration of the simulation. Individual slices through the $\zeta$ distribution across the ejection at $\simeq200$\,ms are shown in Figure \ref{figure:hz}. In contrast, the low luminosity model has intermittent ejections with much less $\zeta>\zeta_{\rm crit}$ material per ejection.}
\label{figure:mass}
\end{figure*}

Figure \ref{figure:mass} shows the mass ejected as a function of $\zeta_r$ for all of the mass in the computational domain in the critical temperature range for nucleosynthesis between $0.2\leq T\leq 0.5$\,MeV. 
For the high luminosity $\lumanue=8\times10^{51}$\,ergs/s model (left panel), we find well-defined periodic ejections separated by $\sim90$\,ms. The right panel shows the low luminosity model with $\lumanue=4\times10^{50}$\,ergs/s, which exhibits much more erratic behavior, and although the ejections come every few hundred milliseconds, we do not observe a strict periodicity over the relatively short time of the simulation.

\begin{figure*}
\centerline{\includegraphics[width=4.3cm]{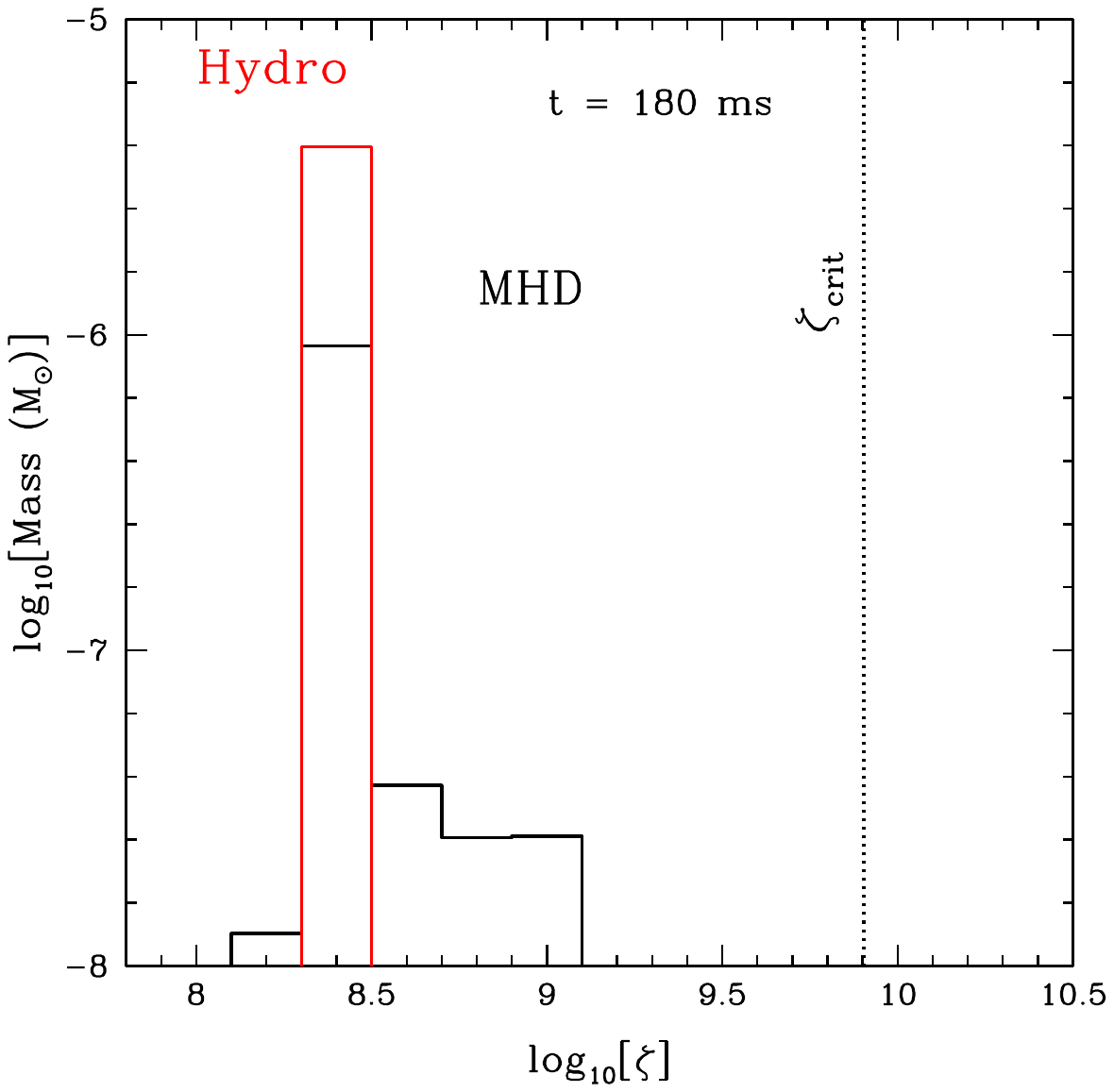}
\includegraphics[width=4.3cm]{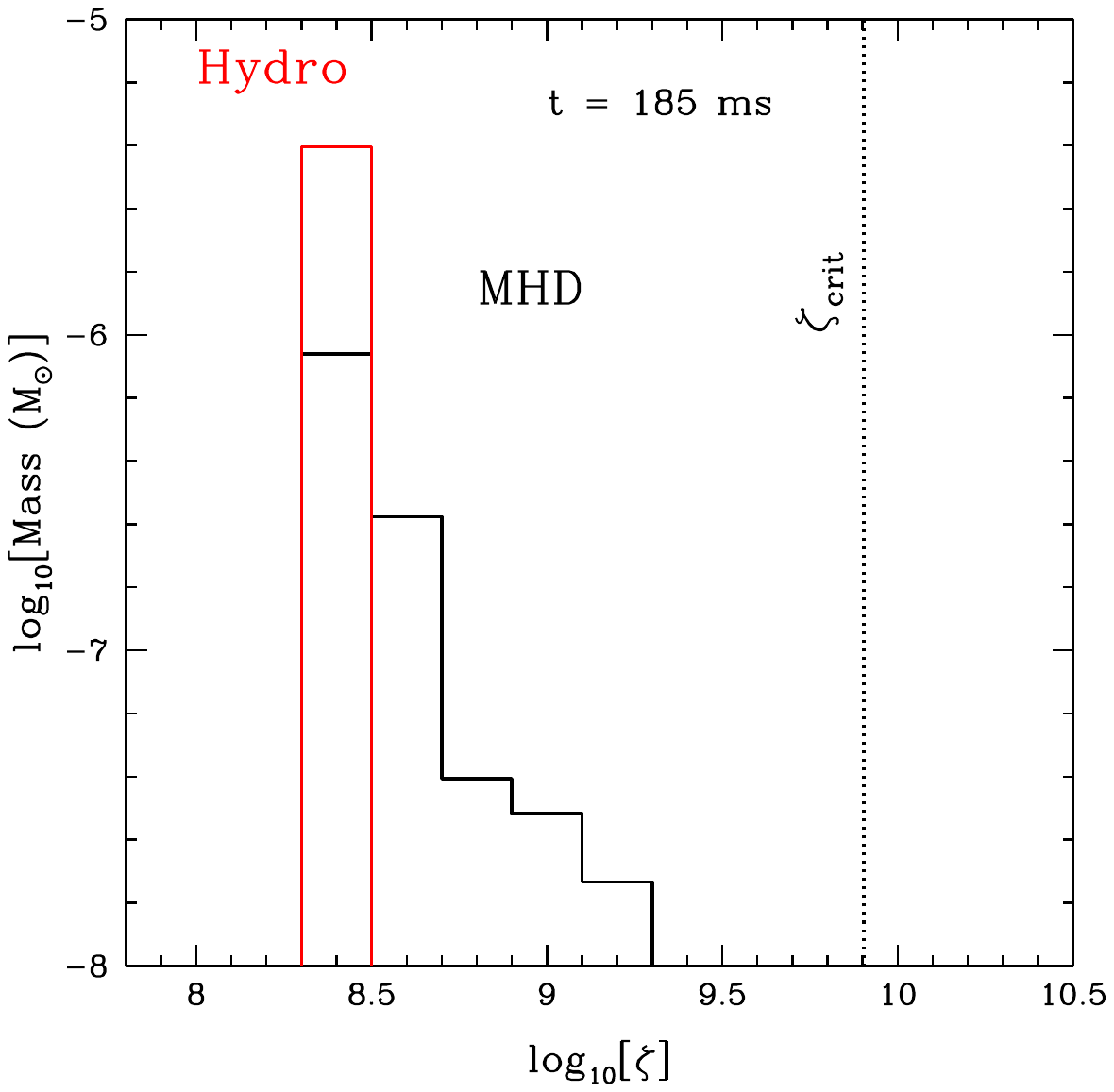}
\includegraphics[width=4.3cm]{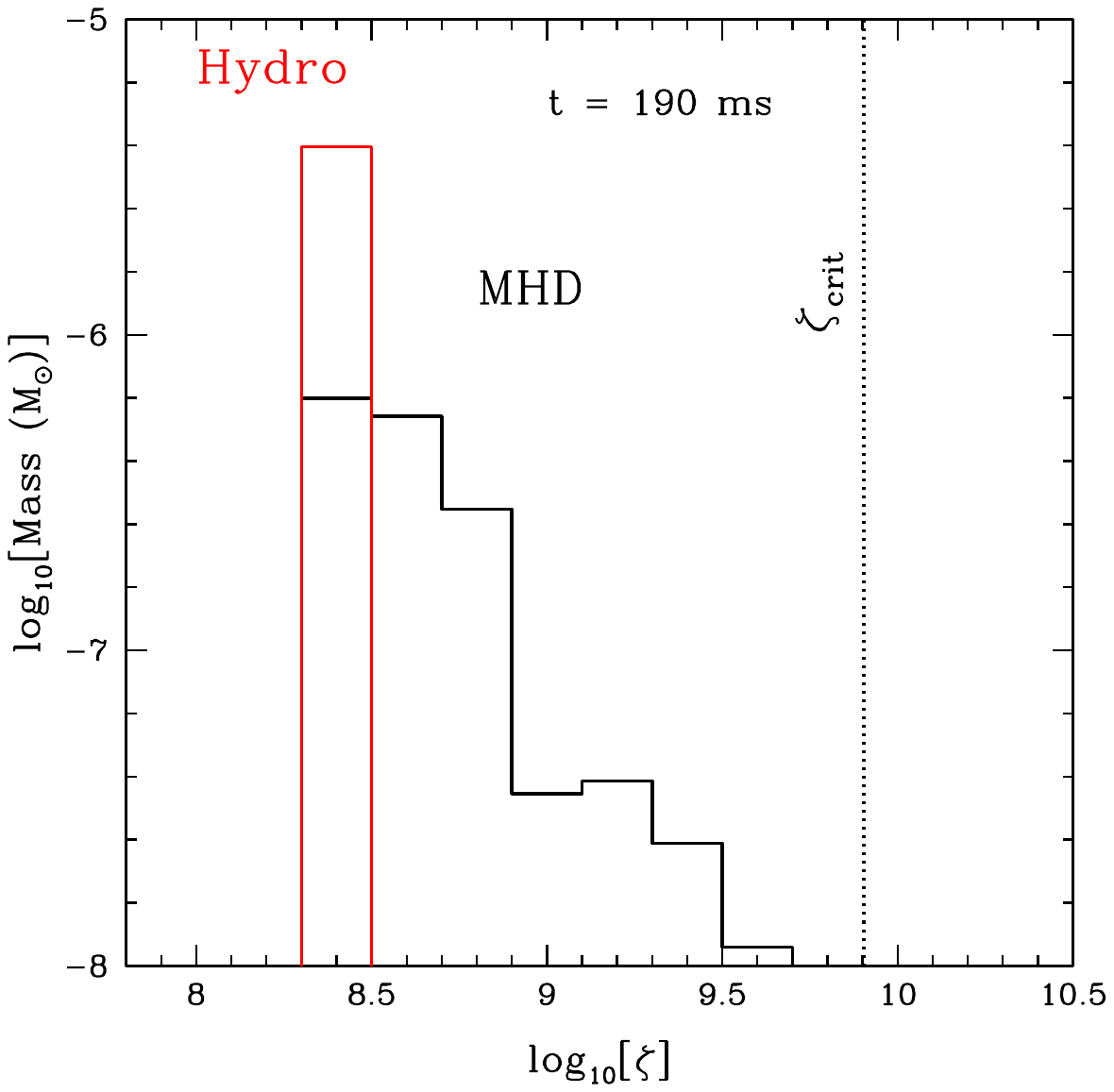}
\includegraphics[width=4.3cm]{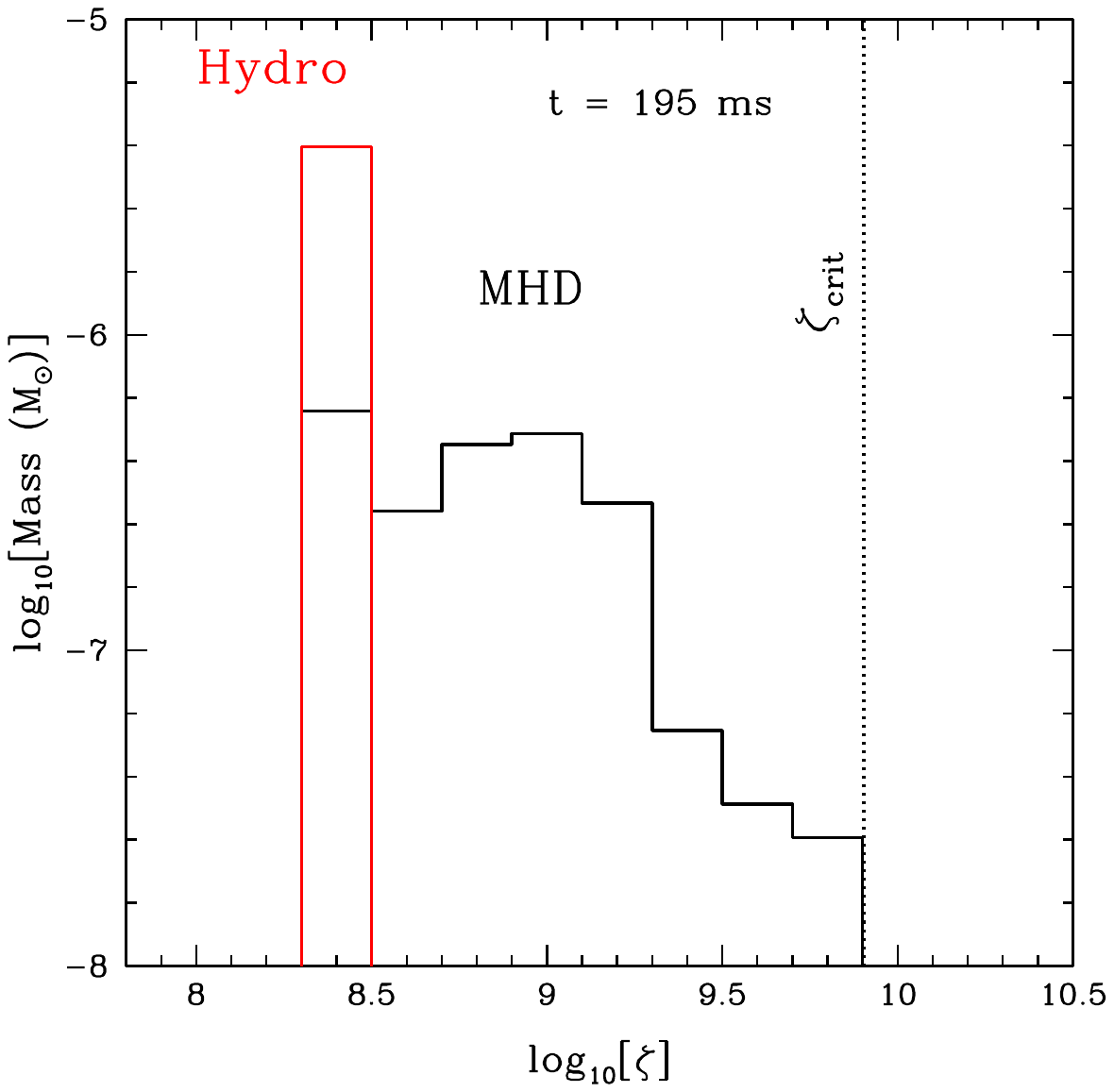}}
\vspace*{-1.5cm}
\centerline{
\includegraphics[width=4.3cm]{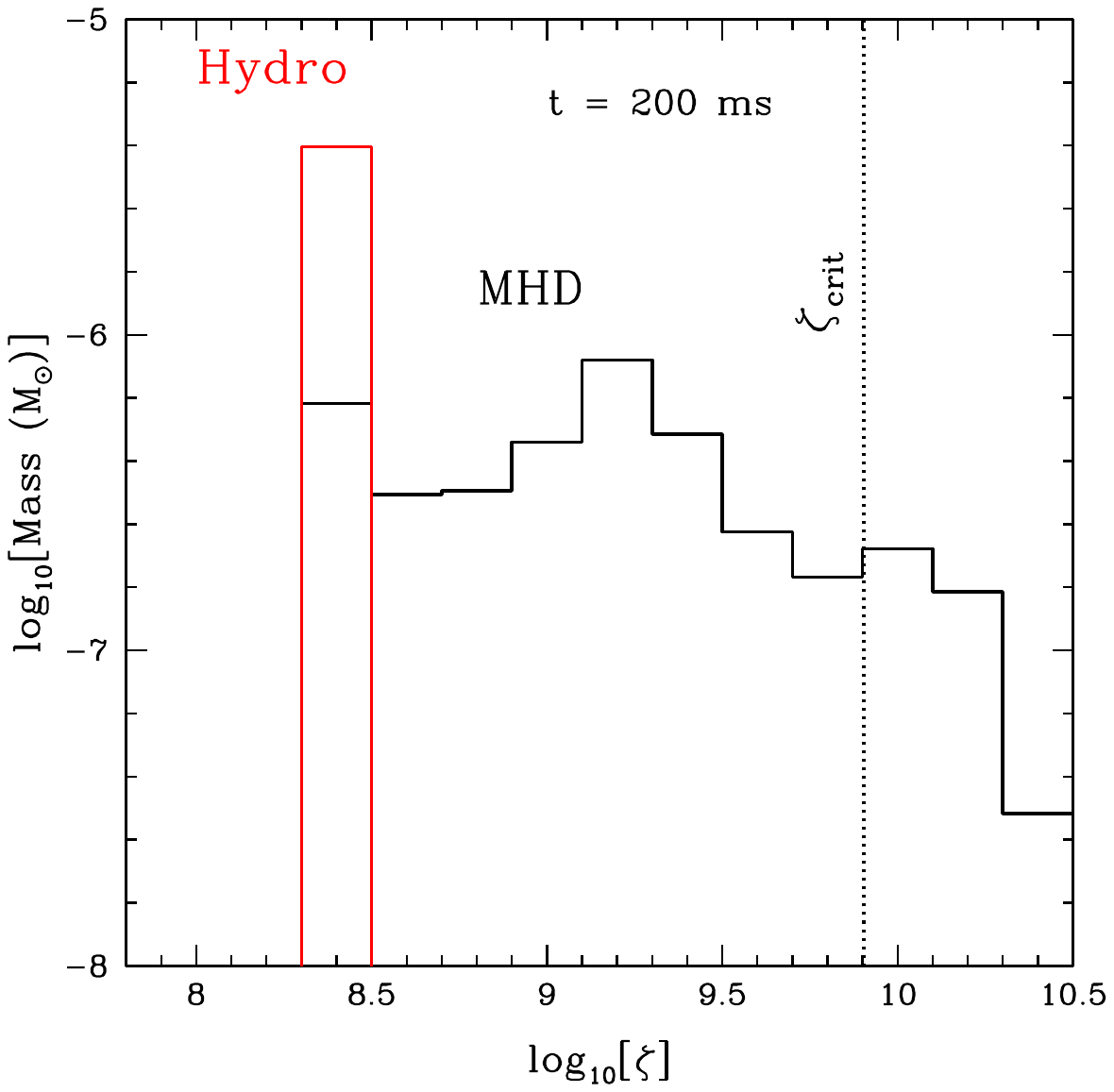}
\includegraphics[width=4.3cm]{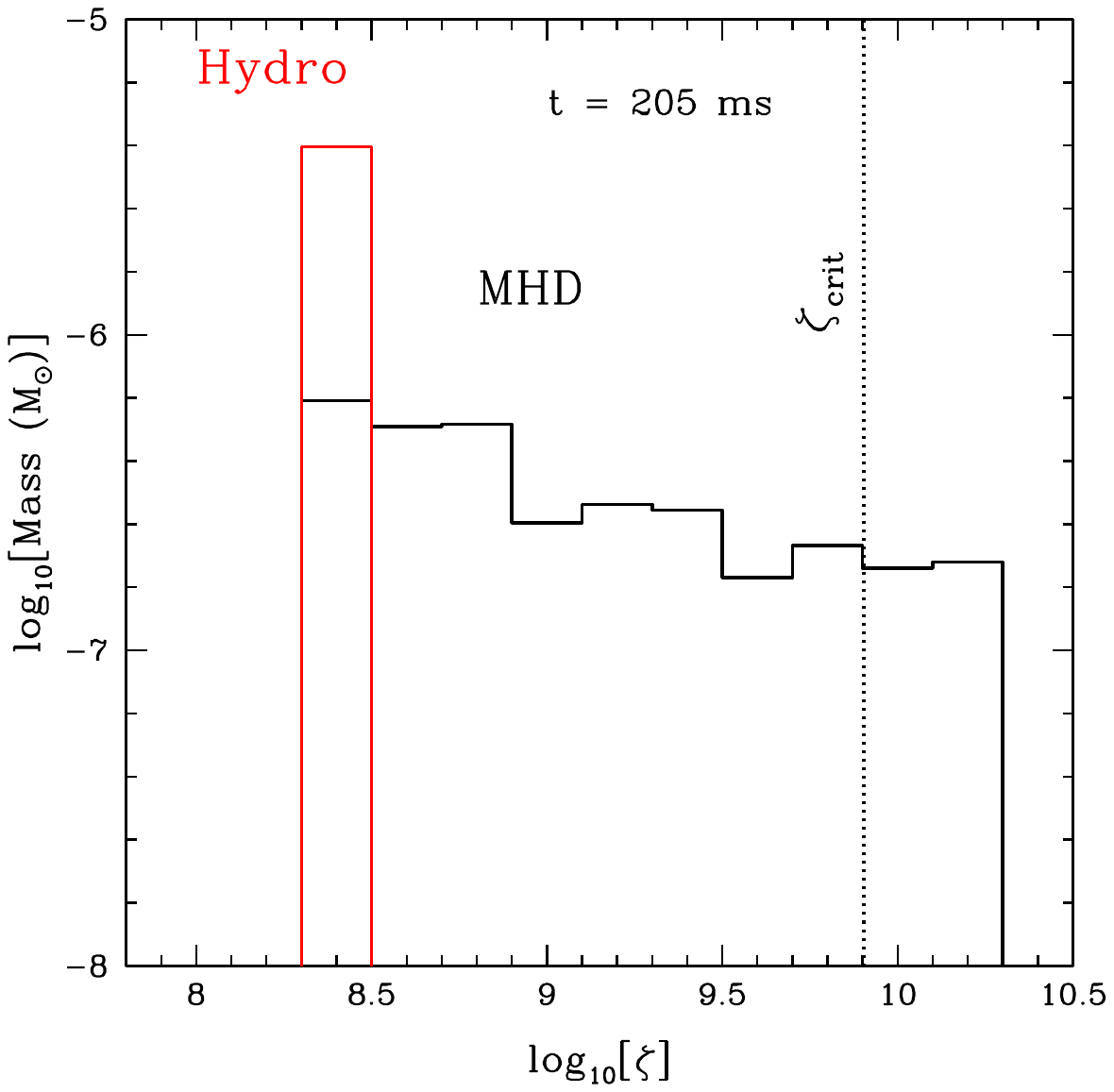}
\includegraphics[width=4.3cm]{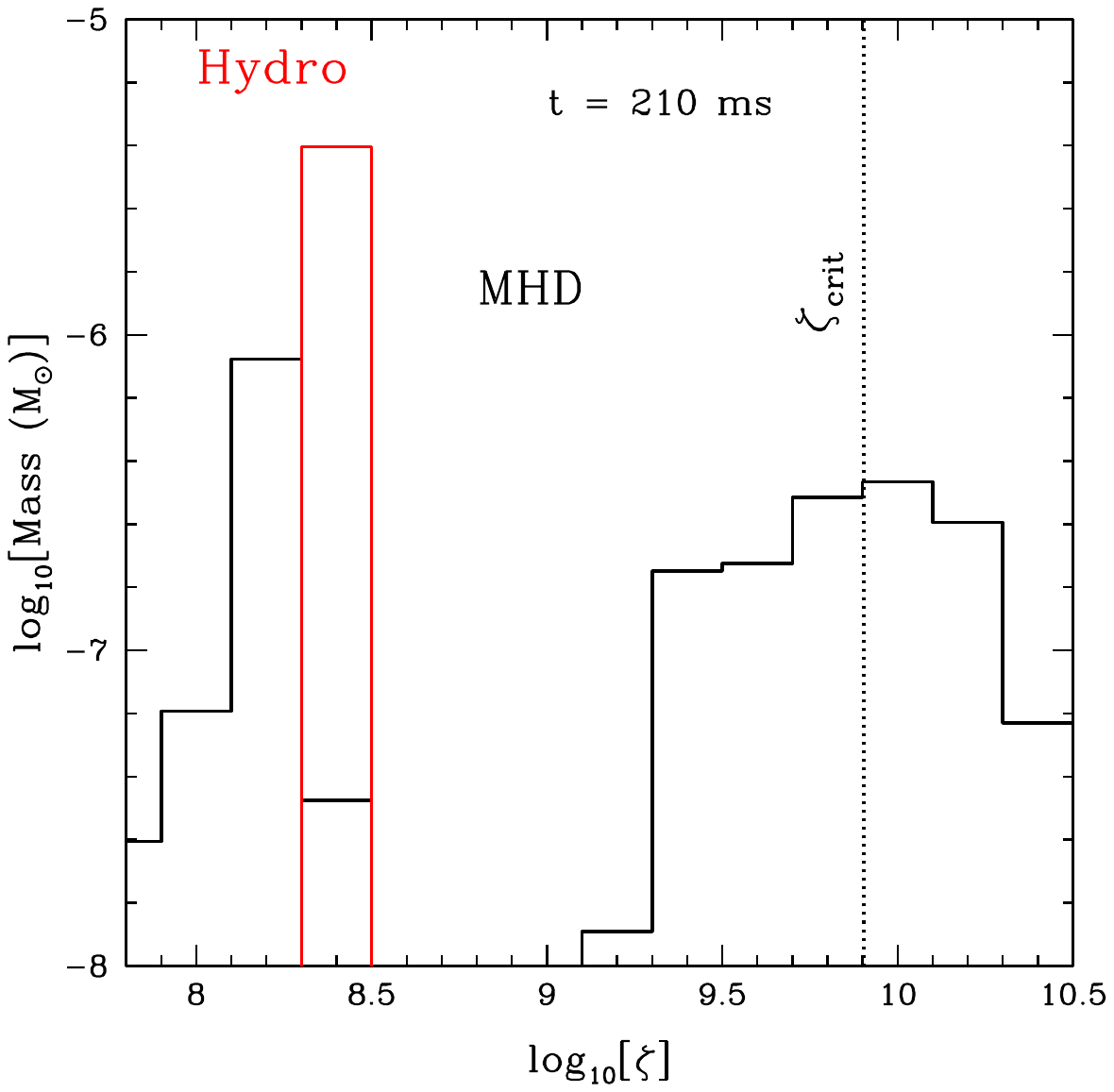}
\includegraphics[width=4.3cm]{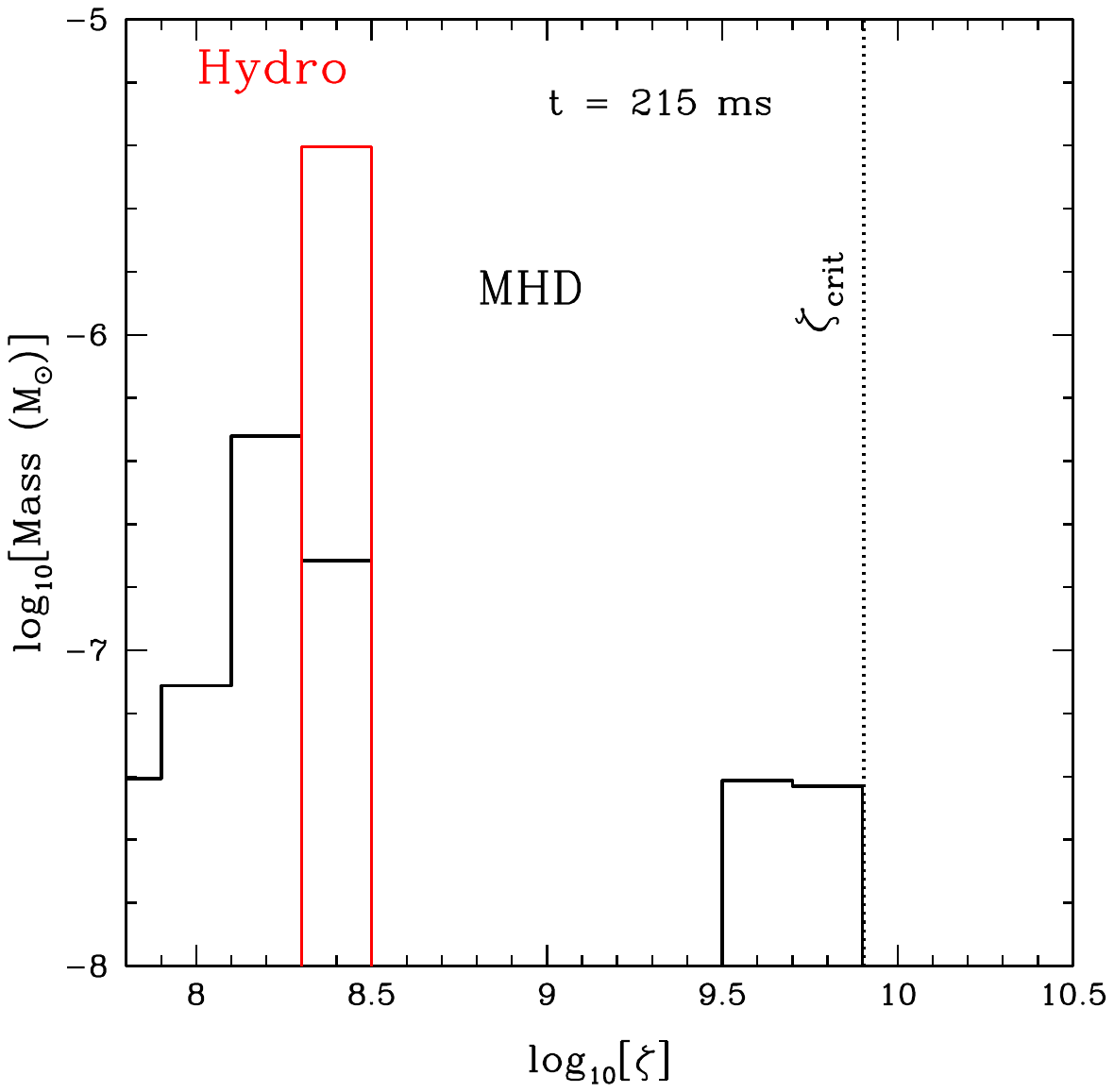}}
\vspace*{-1cm}
\caption{Snapshots of the total mass in the computational domain in the temperature interval $0.2\leq T\leq 0.5$\,MeV as a function of $\log_{10}[\zeta]$ for the high luminosity calculation shown in Figures \ref{figure:highl} and \ref{figure:mass} (left panel) for the plasmoid ejection at $t\sim200$\,ms. Times are separated by 5\,ms and noted in each panel. The results from a purely hydrodynamical calculation $B=0$ with the same neutrino luminosity are shown for comparison (red histogram; same in all panels). The dashed vertical line denotes $\zeta_{\rm crit}$ from equation (\ref{zetacrit}). The upper left panel shows the beginning of plasmoid emergence, while the lower right panel shows after ejection as the magnetosphere's closed zone resets before the next ejection.}
\label{figure:hz}
\end{figure*}

\begin{figure*}
\centerline{\includegraphics[width=9cm]{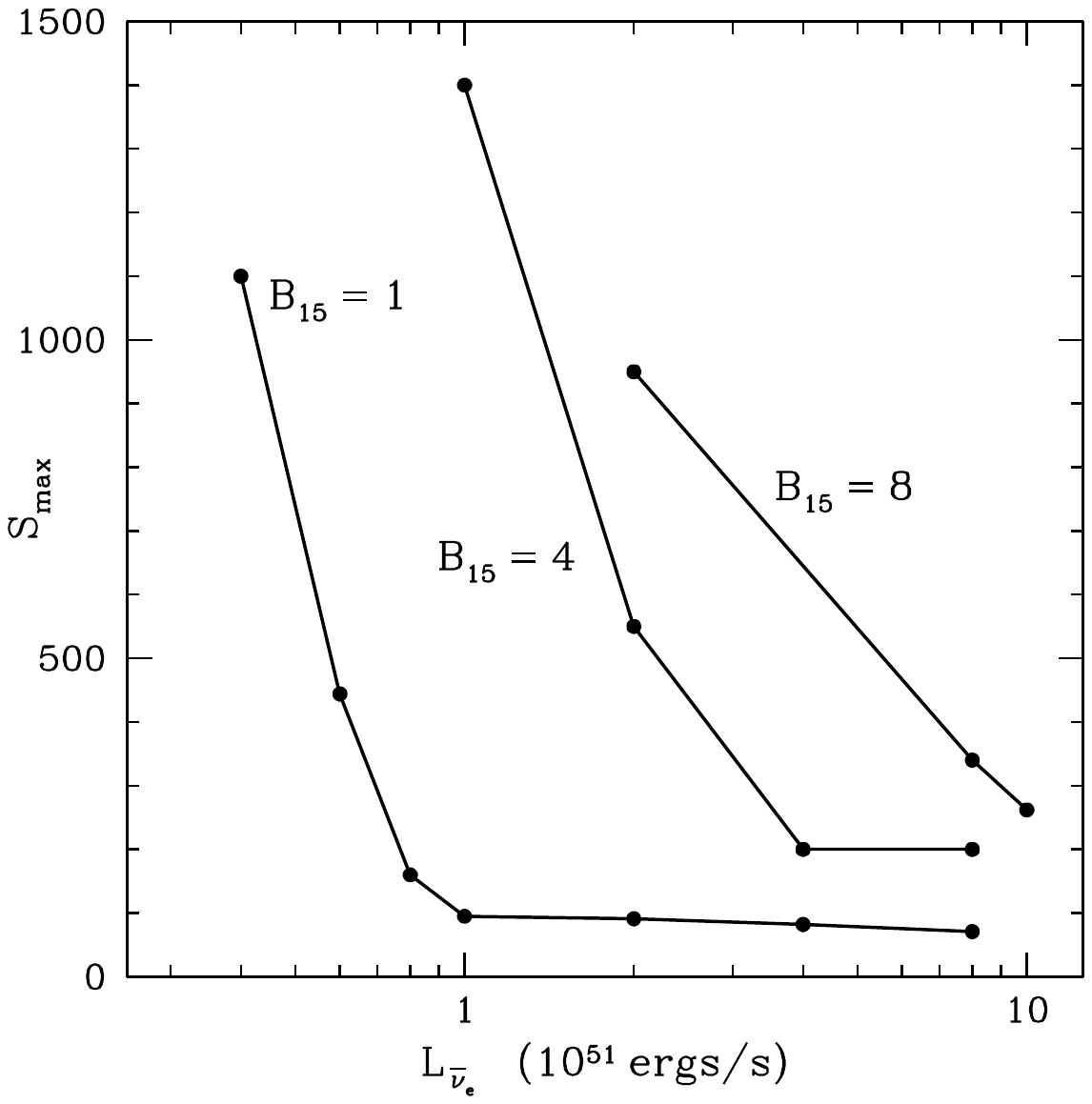}\includegraphics[width=9cm]{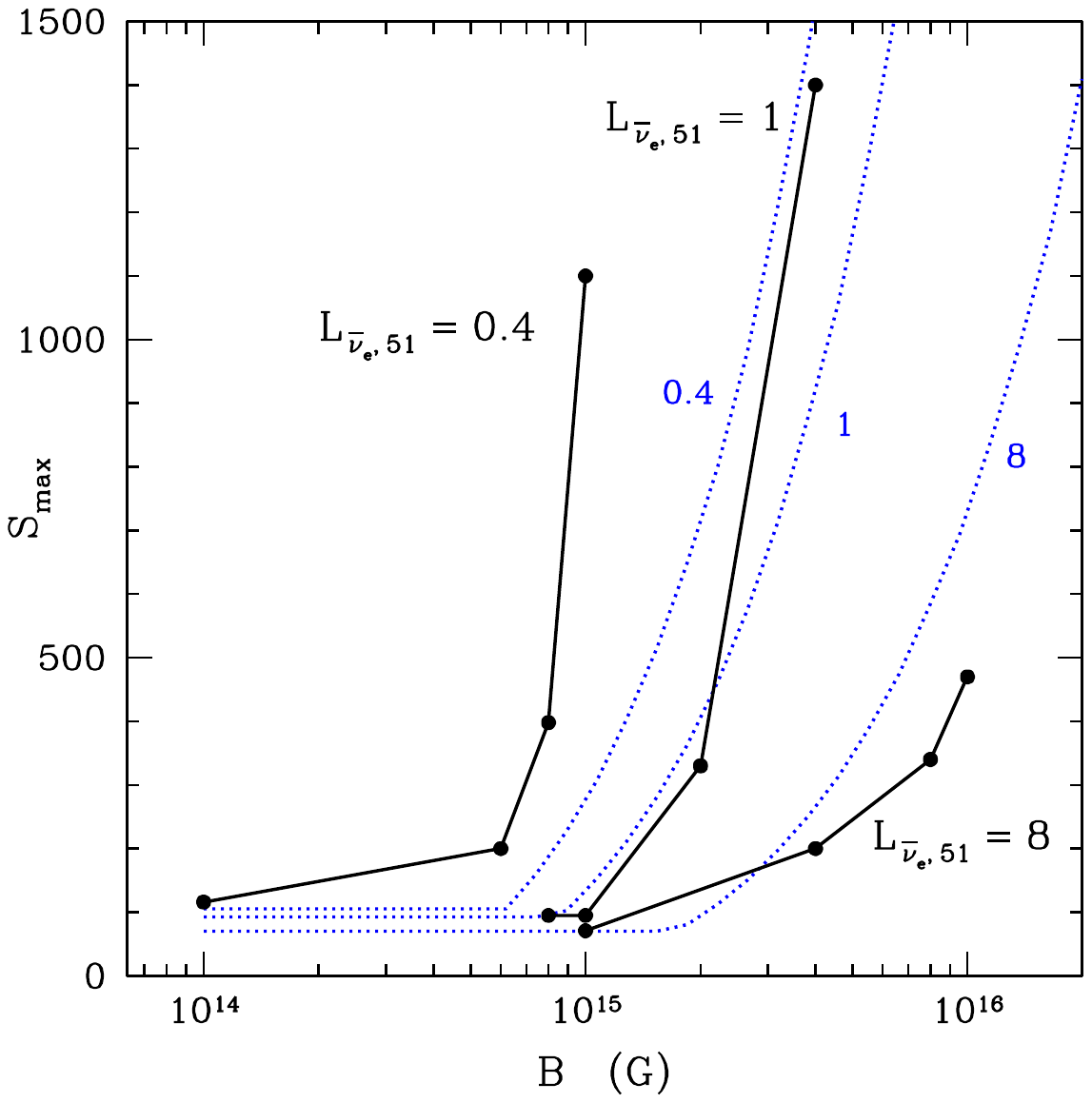}}
\vspace*{-1.8cm}
\caption{Maximum mass-weighted entropy in the erupting plasmoids, averaged over several ejection episodes, as a function of neutrino luminosity for several magnetic field strengths (left), and as a function of magnetic field strength for several values of the neutrino luminosity (right). Analytic predictions as described in Section \ref{section:analytic}  for $L_{\bar{\nu}_e}=0.4$, 1, and $8\times10^{51}$\,ergs s$^{-1}$ (dotted blue lines) are shown in the right panel. }
\label{figure:smax}
\end{figure*}

Figure \ref{figure:hz} shows snapshots of the time evolution of the distribution of mass in the temperature interval $0.2\leq T\leq 0.5$\,MeV as a function of $\zeta$ during a single ejection from the high luminosity model at approximately $t\sim200$\,ms (see left panel of Fig.~\ref{figure:mass}). The solid red histogram shows the result for a purely hydrodynamical model with $B=0$ for comparison, while the solid black lines show snapshots of the evolution separated by $5$\,ms from $t=180$\,ms (upper left) to $t=215$\,ms (lower right), corresponding to before, during, and then after the corresponding plasmoid ejection shown in Figure \ref{figure:mass}. $\zeta_{\rm crit}$ is marked by the vertical dotted line. The MHD models produce a broad distribution of ejected mass as a function of $\zeta$, which will imprint itself in the resulting nucleosynthetic abundances. The total amount of mass above $\zeta_{\rm crit}$ and the expected yield are discussed in Section \ref{section:mass}.

Figure \ref{figure:smax} summarizes results for a wide range of PNS models with different $B$ and $\lumanue$. It shows the maximum entropy, $S_{\rm max}$, obtained by averaging over several ejected plasmoids, as a function of the PNS magnetic field strength (right panel) and neutrino luminosity (left panel). For a given luminosity, $S_{\rm max}$ rapidly increases as a function of magnetic field strength. The critical value of the dipole magnetic field required for an order-unity increase in the maximum entropy is approximately
\beq
B_{\rm crit}\sim10^{15}\,{\rm G}\,L_{\bar{\nu}_e,\,51}^{1/2}\,\,\,(0.1\lesssim L_{\bar{\nu}_e,\,51} \lesssim10)
\label{bcrit}
\eeq
In an evolving PNS wind, we expect the neutrino luminosity to decrease as roughly as $\sim1/t$ from an initial value of $\sim10^{52}$\,ergs s$^{-1}$ for $\sim30-100$\,s \citep{pons1999}. For a given dipole magnetic field strength, we thus expect the field to become increasingly strong relative to the wind, and for the entropy in the ejected plasmoids to rapidly increase with time. In addition, the average magnetic dipole field strength may evolve in time throughout the cooling epoch as a result of dynamo action, complicating the evolution of the entropy and nucleosynthetic yield. An important point is that in order to obtain the high entropy ejections we see here, the magnetosphere must be stable for of order the heating and ejection timescale $\sim100$\,ms (see Section \ref{section:discussion}).

The dynamics we observe in our simulations are apparently generic to magnetically dominated winds with heating. In particular, research in the Solar wind context has shown that the helmet-streamer configuration of \cite{pneuman1971} can be unstable to periodic ejections. \cite{suess1996} first discussed the possibility that the structure could evolve in time. \cite{endeve2003,endeve2004} showed that the structure is unstable to the periodic ejection of matter in a manner qualitatively similar to Figures \ref{figure:highl} and \ref{figure:lowl} when the simulations include volumetric heating and conductive cooling. \cite{endeve2003,endeve2004} also found that the dynamics were sensitive to the cooling efficiency, as measured in their simulations by the coefficient of conductivity (see also \citealt{chen2009,allred2015}); for larger cooling rates, the structure becomes stable in their calculations. As \cite{endeve2003} describe, the combination of heating with the very different pressure scale height between the open and closed regions of the magnetosphere yields periodic ejections. As the magnetosphere is heated, it grows until the pressure dominates magnetic forces, and the plasmoid erupts. After mass expulsion from the closed zone, the magnetosphere reforms via reconnection at the magnetic equator, in some cases effectively dragging wind material back toward the star. The cycle then repeats. The behavior we observe is also qualitatively similar to the results reported by \cite{komissarov_barkov}, who explored the dynamics of the wind and jets launched by millisecond spin period magnetars relevant to GRBs. The rapid rotation in their simulations causes significant magneto-centrifugal acceleration and much higher overall velocities.

\subsection{Mass Ejected and Expected Yield}
\label{section:mass}

Figures \ref{figure:mass} and \ref{figure:hz} show that only a small fraction of the total wind mass loss rate reaches $\zeta>\zeta_{\rm crit}$. The $\lumanue=8\times10^{51}$\,ergs/s model ejects $M(\zeta\gtrsim\zeta_{\rm crit})\sim7\times10^{-7}$\,M$_\odot$ per eruptive event every $\sim90$\,ms, implying a total mass outflow rate in high-$\zeta$ plasmoids of $\dot{M}(\zeta\gtrsim\zeta_{\rm crit})\sim8\times10^{-6}$\,M$_\odot$/s. This is $\simeq8$\% of the total mass loss rate of $\dot{M}_{\rm tot}\simeq1\times10^{-4}$\,M$_\odot$/s. Since the timescale for the PNS luminosity to decrease at these high luminosities is of order $\sim1$\,s, we expect $\sim8\times10^{-6}$\,M$_\odot$ of high-$\zeta$ material to be ejected just after the wind begins, at high neutrino luminosities. 

For the low luminosity $\lumanue=4\times10^{50}$\,ergs/s model, the total mass loss rate is $\dot{M}_{\rm tot}\simeq2.4\times10^{-7}$\,M$_\odot$/s, with $M(\zeta\gtrsim\zeta_{\rm crit})\sim1-2\times10^{-8}$\,M$_\odot$ per ejection and an (intermittent) ejection timescale of $\sim300$\,ms, implying a mass-loss rate in high-$\zeta$ material of $\dot{M}(\zeta\gtrsim\zeta_{\rm crit})\sim3-6\times10^{-8}$\,M$_\odot$/s. Assuming that similar behavior continues for the timescale for the PNS luminosity to change at low luminosity $\sim10$\,s implies roughly $\sim3-6\times10^{-7}$\,M$_\odot$ in high-$\zeta$ material ejected. 

Taking into account our calculations of the mass loss rates at other neutrino luminosities (see Fig.~\ref{figure:smax}), these estimates imply that of order $\sim10^{-5}$\,M$_\odot$ of high-$\zeta$ material may be ejected throughout the cooling epoch, but with a strong dependence on $B$. For fields as high as $B\sim10^{16}$\,G, the wind becomes magnetically-dominated just as the explosion is occurring. Indeed, for high magnetic fields of $\gtrsim10^{16}$\,G, our calculations may underestimate the total ejection of high-$\zeta$ material because $\sim1-{\rm few}$ massive eruptive events may occur during the explosion itself, just as the wind phase is beginning. In contrast, for $B\sim10^{15}$\,G, the field does not dominate the dynamics until late into the cooling epoch, with lower neutrino luminosities, and lower mass loss rates overall and we expect $\sim10$ times smaller total mass in high-$\zeta$ ejecta, $\sim10^{-6}$\,M$_\odot$.

Based on our calculations of $\zeta$ during the eruptions, if the wind is neutron-rich we expect the high-$\zeta$ material to produce an $r$-process that extends to the actinides. However, the fraction of the matter with $\zeta>\zeta_{\rm crit}$ that ends up as heavy elements is strongly dependent on $Y_e$. \cite{hoffman97} give final $\alpha$ particle fractions as a function of $Y_e$, $S$, and $t_{\rm dyn}$ for a series of models with $\zeta\simeq\zeta_{\rm crit}$ (their Table 5). This work implies that for $Y_e\gtrsim0.45$, as expected from PNS cooling calculations (e.g., Fig.~4 from \citealt{vlasov2017}), the yield of heavy elements is $1-2Y_e\sim0.1$. These $\alpha$-rich freezeouts from the preceding $\alpha$-process mean that if $\sim10^{-5}$\,M$_\odot$ of high-$\zeta$ material is ejected throughout the cooling epoch, $\sim10^{-6}$\,M$_\odot$ of heavy elements will be produced. As we discuss in Section \ref{section:discussion}, this yield of heavy $r$-process elements per magnetar birth is prima facie too low to explain the total budget of heavy $r$-process elements in the Galaxy, but the overall assessment of magnetars as an $r$-process production site must await dynamical calculations including rotation, and following the evolution from explosion through the cooling epoch. If a single massive eruptive event dominates the production of heavy elements it may be necessary to evolve from a self-consistent MHD supernova calculation.

The high-latitude mass ejected in the steady wind has $\zeta<\zeta_{\rm crit}$ and should produce a ``weak" $r$-process that extends to the first and second abundance peaks, as in the static magnetic field magnetar calculations of \cite{vlasov,vlasov2017}. In our high luminosity calculations shown in Figures \ref{figure:highl}, \ref{figure:mass}, and \ref{figure:hz}, only $\sim$\,5-10\% of the ejected material has $\zeta>\zeta_{\rm crit}$, while the remaining $\sim$\,$90-95$\% has $\zeta<\zeta_{\rm crit}$. However, our calculations show that there is significant time-dependent modulation of $dM/d\zeta$ (see Figs. \ref{figure:mass} and \ref{figure:hz}) as a result of the dynamics of the opening and closing magnetosphere. Each eruption produces a broad distribution of thermodynamic trajectories. In fact, after each eruption, the mass weighted $\zeta$ distribution drops to {\it lower} values than the purely hydrodynamical calculation (lower right panels in Fig.~\ref{figure:hz}), mostly as a result of the very long dynamical timescales for the material that becomes trapped as the magnetosphere closes. It remains to be understood how this broad time-dependent distribution of $\zeta$ imprints itself on the final abundance pattern. In particular, whether or not $r$-process or $p$-process nuclei are produced, the high-entropy ejections we find may produce unique abundance patterns.

\subsection{Observational Implications: Chemical Enrichment \& Binary Pollution}
\label{section:observations}

Although the overall yield of heavy elements from the time-integrated evolution of a cooling highly-magnetic neutron star is uncertain, it is worth asking how this production mechanism might impact observations. In particular, the systems discussed in this paper would provide a prompt heavy element enrichment channel in the universe that should track early Fe production.

We assume that a mass of high-$\zeta$ material $M_\zeta$ is ejected per highly-magnetized NS birth and that a fraction $f_H$ produces heavy elements ($M_H = f_H M_\zeta$; $f_H\simeq1-2Y_e\sim0.1$ for the $r$-process) with mass numbers $A$, and mass of a given element of $M_A=f_A f_H M_\zeta$.  A typical core-collapse supernova produces a mass of iron of $M_{\rm Fe}\simeq0.03$\,M$_\odot$, implying that
\beq
\frac{M_A}{M_{\rm Fe}}=\frac{f_A f_H M_\zeta}{M_{\rm Fe}} 
\simeq3\times10^{-7}f_{A,-2} f_{H,-1}\,\frac{M_{\zeta,\,-5}}{M_{\rm Fe\,-1.5}},
\label{fa}
\eeq
where we have scaled to a yield in an individual heavy element (all isotopes) of $f_{A,-2}=f_A/0.01$, with a heavy element production fraction of $f_{H,-1}=f_H/0.1$, assuming the $\alpha$-rich freezeouts discussed in Section \ref{section:mass}, and where $M_{-x}=M/10^{-x}$\,$M_\odot$. For a commonly observed $r$-process element like Eu, this translates to $\left[{\rm Eu/\rm Fe}\right]\simeq0$ for $f_{\rm Eu}=0.01$, or approximately the Solar ratio \citep{Lodders}. If the supernova ejects $\sim10$\,M$_\odot$ of H, we find that $\left[{\rm Eu/\rm H}\right]\simeq0.3$ for the ejecta, for the parameters above. If highly-magnetized neutron stars form in an early epoch in the evolution of the universe, they should thus provide prompt heavy element enrichment.

Binary companions may also be polluted when the primary experiences a supernova. We assume a long-lived binary companion of $\sim1$\,M$_\odot$ at distance $D$ so that the geometrical cross section implies that a fraction $f\sim R^2/4D^2\sim5\times10^{-6}(R/R_\odot/{\rm AU}/a)^2$ of the ejected mass may pollute the companion if it is ejected into $4\pi$, where the semi-major axis of the orbit $a$ has been scaled to 1\,AU and where we assume a main sequence dwarf with $R=R_\odot$ at the time of the supernova. The total mass deposited in high-$\zeta$ material would then be $\sim5\times10^{-11}(M_\zeta/10^{-5}\,M_\odot)(R/R_\odot/{\rm AU}/a)^2$\,M$_\odot$.  Interestingly, such a model for binary pollution would predict a direct correlation between the amount of iron-peak and heavy element enrichment since, naively, the ratio of the heavy element mass to the iron-peak mass intercepted by the binary companion is a constant as a function of semi-major axis. As an example for simplicity, if $10^{-3}=f_Af_H$ of the mass of high-$\zeta$ ejected material forms Eu, and $0.03$\,M$_\odot$ is ejected in Fe, then the ratio would simply be $\left[{\rm Eu/\rm Fe}\right]\simeq0$, as in the estimate above. If the companion starts with zero metallicity, we can then estimate the iron and Eu content that would be observed.  The material deposited during the primary's supernova will be mixed into the mass of the star's convective layer. As the star evolves, the convective mass grows. For a red giant, as observed in ultra-faint dwarf galaxies like Reticulum II (e.g., \citealt{Ji2016}) and in the Galactic halo, the convective mass is of order $M_{\rm con}\simeq0.4$\,M$_\odot$, implying that 
\beq
\left[\frac{{\rm Eu}}{\rm H}\right]=\log\left[\frac{M_{\rm Eu}f}{M_{\rm con} X_{\rm H} \langle A\rangle}\right]+11.48\sim-3.5, 
\eeq
where $M_{\rm Eu}\sim10^{-8}(M_\zeta/10^{-5}\,M_\odot)\,f_{\rm Eu,-2}\,f_{\rm H,-1}$\,M$_\odot$ is the total synthesized Eu mass, $f$ is the fractional cross section of a dwarf star at 1\,AU, $X_{\rm H}=0.71$, $\langle A\rangle\simeq152$ for Eu, and $-11.48=0.52-12$ is the Solar Eu abundance relative to Hydrogen \citep{Lodders}. The iron abundance of the companion would be similar: $\left[{\rm Fe/\rm H}\right]\simeq-3.5$. In such a picture, the range of $[{\rm Eu/H}]$ and $[{\rm Fe/H}]$ observed in a sample of stars would be interpreted simply as different semi-major axes for the companions at the time of explosion, scaling as $a^{-2}$. If the polluted companion were observed as a dwarf, the relative heavy element abundance would increase by a factor of the ratio of the convective mass, $\sim0.4/0.02\sim20$. 

Although these numbers do not work for explaining the Reticulum II giant abundances from individual binary pollution events --- $\left[{\rm Eu/\rm Fe}\right]\sim1.7$ is observed, much higher than estimated above --- these simple scalings imply that the individual abundances of heavy elements like Eu may be observable from stars whose companions explode as supernovae. Moreover, the ejecta in supernovae is highly velocity stratified, with material extending from the usual $\sim5000$\,km/s to just $\sim100$\,km/s and below \citep{kifonidis}. The slower ejecta may be more easily captured by a binary companion. In particular, if the material's velocity $V$ is slower than the escape velocity of the binary companion $V_{\rm esc}$, as expected for the most inner ejecta of the explosion, the cross section for capture could be increased by $\sim(V_{\rm esc}/V)^2$ (for a recent related discussion, see, e.g., \citealt{liu2015}). For a M$_\odot$ dwarf, $\sim(V_{\rm esc}/V)^2\sim40$ for $\sim100$\,km/s ejecta, implying a big overall boost to the heavy element pollution, but also the potential for differential pollution between between elements that might have different velocity distributions (e.g., Eu versus Fe or O).

\subsection{Semi-Analytic Model for Plasmoid Eruption}
\label{section:analytic}

Following \cite{thompson03}, we estimate the maximum entropy amplification and trapping timescale for the plasmoids in a simple one-zone model in an effort to provide some explanation for the behavior we find in our simulations.

We first ask at what polar dipole magnetic field strength we expect the magnetic tension force to dominate the wind dynamics. Consider a freely expanding unmagnetized wind pressure, density, velocity, and temperature profile for a given neutrino luminosity and PNS mass and radius. For free spherical winds, the sonic point is at a radial location of $\sim300-3000$\,km \citep{tbm}, and the thermal pressure profiles in the inner region near the PNS $r\lesssim100$\,km are thus well approximated by hydrostatic equilibrium. As a consequence, the thermal pressure is much greater than the kinetic energy density of the outflow ($P>\rho V^2/2$) near the PNS. To estimate the radial range over which the magnetic field dominates the flow, we can therefore compare the magnetic energy density associated with the magnetic tension force in the equatorial plane with the thermal pressure profile of the wind without magnetic forces. 

Take a simple dipole magnetic field with surface polar field strength $B$. The magnetic energy density at the equator is
\beq
\frac{B^2_{\rm eq}(r)}{8\pi}=\frac{1}{8\pi}\left(\frac{B}{2}\right)^2\left(\frac{R_{\nu}}{r}\right)^6.
\eeq
An order-of-magnitude estimate for the energy density associated with the magnetic tension force in the equatorial plane is then
\beq
u_{\rm B}\sim\frac{B^2_{\rm eq}(r)}{4\pi}\left(\frac{R_\nu}{R_c}\right)
\label{tension}
\eeq
where $R_c$ is the radius of curvature of the magnetic field.

\begin{figure*}
\centerline{\includegraphics[width=9cm]{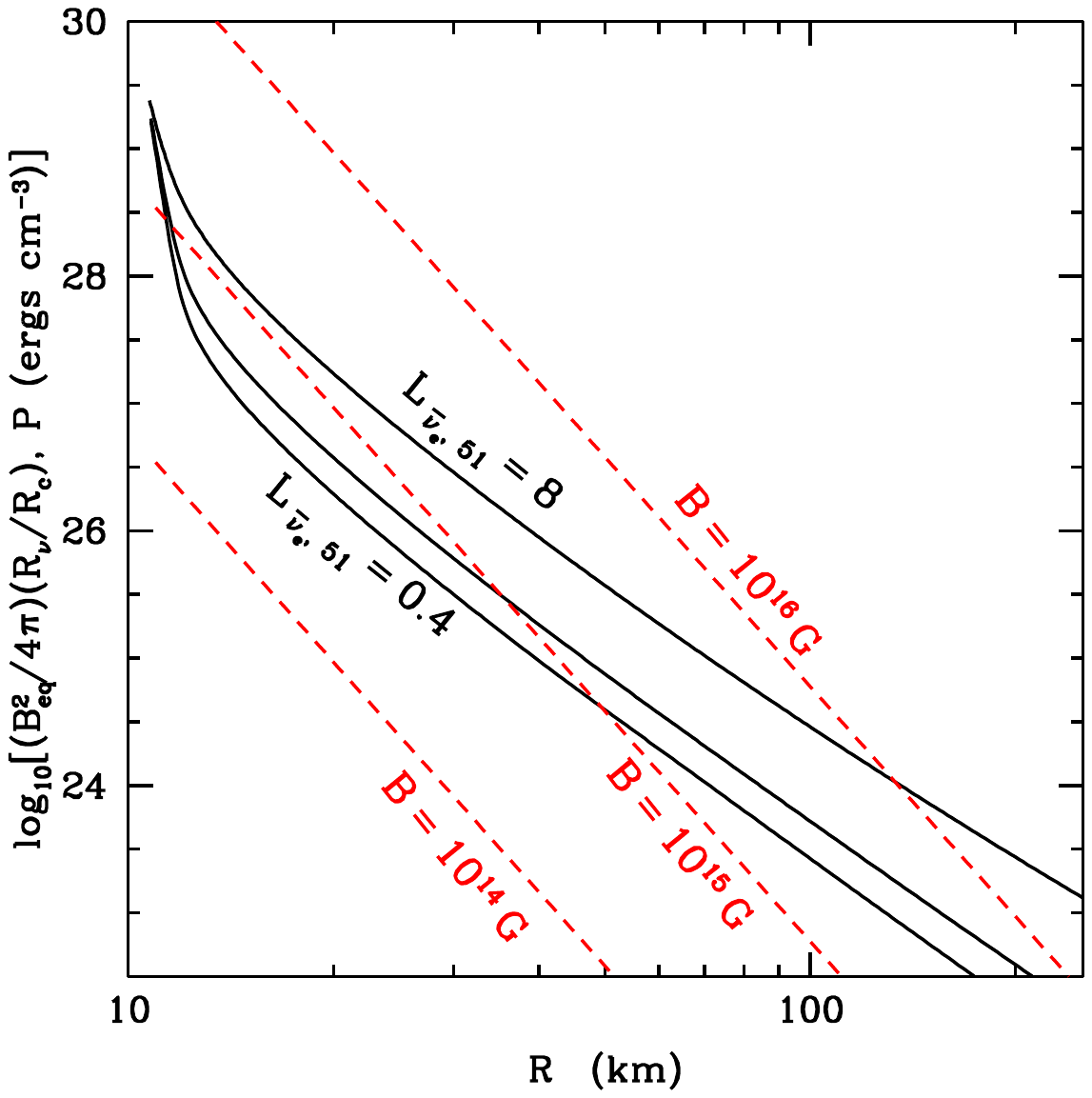}\includegraphics[width=9cm]{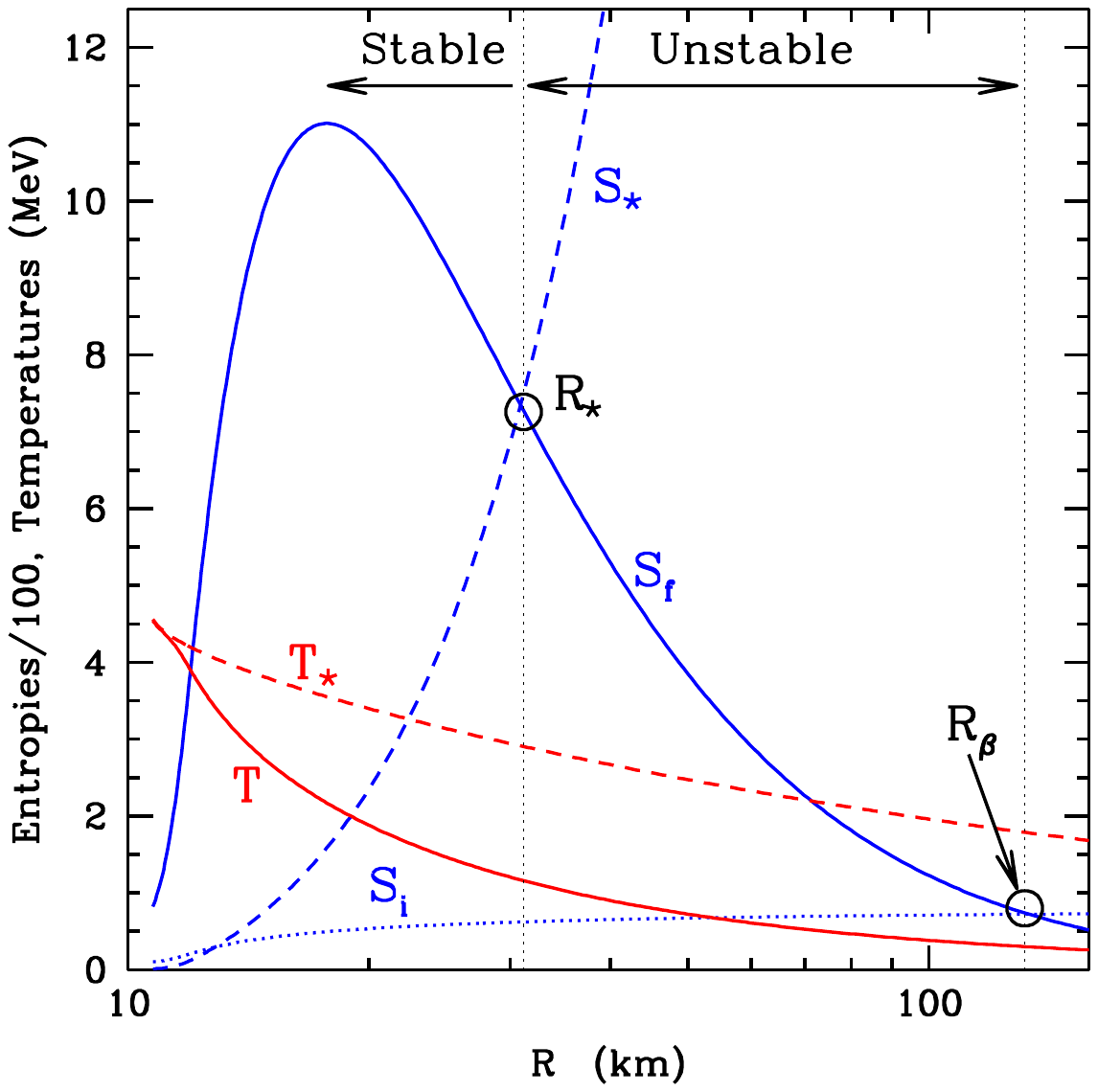}}
\vspace*{-1.8cm}
\caption{{\it Left:} Thermal pressure profiles for un-magnetized spherical winds with $\lumanue=8$, 1, and $0.4\times10^{51}$\,ergs/s (solid black lines, top to bottom) compared with the estimate of the energy density associated with the magnetic tension force in the equatorial plane given in equation (\ref{tension}) with radius of curvature $R_c=R_\nu/2$ and polar dipole magnetic field strength $B=10^{16}$, $10^{15}$, and $10^{14}$\,G (red dashed lines). {\it Right:} For the specific case of $\lumanue = 8\times10^{51}$\,ergs/s and $B=10^{16}$\,G (e.g., Figs.~\ref{figure:highl}, \ref{figure:mass}, \ref{figure:hz}), as a function of radius from the PNS we show the initial entropy of a un-magnetized wind $S_i$ (dotted blue), final entropy after trapping and enhancement $S_f$ (solid blue) using the analytic expression of equation~\ref{samp}), $S_\star$ (dashed blue; see eq.~\ref{sstar}), un-magnetized wind temperature $T$ (solid red), $T_\star$ (red dashed; see eq.~\ref{tstar}). For $r<R_\beta$, we expect magnetic trapping. For $r<R_\star$ we expect the  closed magnetosphere to remain stable because  $\dot{q}\rightarrow0$  (i.e., $T\rightarrow T_\star$ and $S\rightarrow S_\star$; see Section \ref{section:analytic}). The region $R_\star<r<R_\beta$ should be dynamically ejected on the timescale predicted in equation (\ref{tej}). Predictions for the maximum value of the ejected entropy ($S_f(R_\star)$) for different neutrino luminosities and magnetic field strengths are shown in the right panel of Figure \ref{figure:smax} (blue dotted lines).}
\label{figure:pressure}
\end{figure*}

As an example, in the left panel of Figure \ref{figure:pressure} we show the free wind pressure profile for $\lumanue=8$, 1, and $0.4\times10^{51}$\,ergs s$^{-1}$ (solid black lines) and equation (\ref{tension}) as a function of radius for polar field strengths of $B=0.1$, 1, and $10\times10^{15}$\,G, where we have used a radius of curvature $R_c=R_\nu/2$ for illustration. At and just outside the neutrinosphere $R_\nu$, the pressure is dominated by non-relativistic free nucleons, and drops exponentially with a pressure scale height of a fraction of a kilometer. On somewhat larger scales ($\sim12-15$\,km), as the wind accelerates and the pressure becomes dominated by relativistic electron/positron pairs, the pressure decreases more slowly with radius. As a result, $u_{\rm B}$ falls more rapidly with radius than $P$ on scales larger than the initial exponential atmosphere. Therefore, if $P>u_{\rm B}$ at radii near the PNS surface, the magnetic field will not dominate the flow anywhere, but if $P<u_{\rm B}$, outside the exponential atmosphere, there is a range of radii where $u_{\rm B}>P$ stretching from $\sim R_\nu$ to the radius where $P=u_{\rm B}$, which we define as $R_\beta$.  Looking at the left panel of Figure \ref{figure:pressure}, we see that for $B\lesssim1\times10^{15}$\,G the magnetic field cannot dominate the wind with $\lumanue=8\times10^{51}$\,ergs/s because $P>u_B$ throughout the profile, while for $B$ ranging from $1-10\times10^{15}$\,G $R_\beta$ moves steadily outward to $R_\beta\sim100$\,km. For lower neutrino luminosities, we see that lower $B$ is required to dominate the dynamics near the PNS, but that $B=10^{14}$\,G should not dominate the dynamics of any of the three luminosities plotted here.

For a given $B$, we assume that if $P>u_{\rm B}$, there is no trapping and the magnetosphere is fully opened. Conversely, if $P<u_{\rm B}$ there is magnetic trapping, with $R_{\beta}$ being the outer edge of the magnetosphere. For $r>R_\beta$ we expect free wind conditions to obtain, while for $r<R_\beta$ the material is initially trapped by the magnetic tension force. In the trapped region, the entropy will change according to 
\beq
T\frac{dS}{dt}=\dot{q}
\eeq
where $\dot{q}$ is the net heating rate per gram. A key piece of physics in the PNS wind context is that the neutrino heating and cooling rates are dominated by the charged-current reactions $e^-+p\leftrightarrow n +\nu_e$ and $e^++n\leftrightarrow p+\bar{\nu}_e$, which can be written as \citep{qian_woosley}
\beqa
H_\nu&\simeq& 5.8\times10^{24}f(r)\,R_{\nu,\,10}^{-2}\,\,\, {\rm MeV\,\,\,g^{-1}\,\,\,s^{-1}\,}\nonumber \\
&\times&\left[L_{\nu_e,\,51}\langle\epsilon_{\nu_e, \,\rm MeV}^2\rangle X_n+L_{\bar{\nu}_e,\,51}\langle\epsilon_{\bar{\nu}_e,\, \rm MeV}^2\rangle X_p\right]
\label{h}
\eeqa
where $R_{\nu,\,10}=R_\nu/10$\,km, $L_{\nu_e,\,51}=L_{\nu_e}/10^{51}$\,ergs/s, $X_n$ and $X_p$ are the neutron and proton fractions, respectively,
\beq
f(r)=\left[1-\left(1-\left(R_\nu/r\right)^2\right)^{1/2}\right],
\label{fofr}
\eeq
and the averages of neutrino energies in equation (\ref{h}) are defined as in \cite{tbm} and \cite{qian_woosley}. The net cooling rate is 
\beq
C_\nu=1.37\times10^{24}\,\,\,{\rm MeV\,\,g^{-1}\,\,s^{-1}\,}T_{\rm MeV}^6.
\eeq
Because the neutrino heating rate falls off with radius roughly as $H_\nu\propto r^{-2}$, whereas the cooling rate drops much more rapidly with radius $C_\nu\propto T(r)^6$, heating dominates cooling. For this reason, one generically expects $dS/dt>0$ in the trapped material with $r<R_\beta$. For radii larger than the exponential atmosphere on $\simeq10$\,km scales where the pressure is dominated by relativistic electron/positron pairs, the entropy can approximated by \citep{qian_woosley}
\beq
S\simeq 5.21 T_{\rm MeV}^3/\rho_8
\label{s}
\eeq
where $\rho_8=\rho/10^8$\,g cm$^{-3}$.

As $S$ in the trapped material increases with time, the temperature will increase and the magnetosphere will dynamically adjust. A significant limitation of \cite{thompson03} and the estimates presented here is that we assume the density profile is fixed as the entropy increases. In fact, the temperature and density gradients in the magnetosphere will be set by magnetohydrostatic equilibrium as $S$ increases. Nevertheless, as we show, this approximation provides a useful simplified model for interpreting the simulations.

Once the matter is trapped in the closed zone and the entropy begins increasing, there are two possibilities. The first is that the pressure of the trapped matter increases such that $P$ becomes larger than $u_{\rm B}$. If this condition is met, we expect dynamical ejection. The second possibility is that the temperature increases sufficiently that $C_\nu\rightarrow H_\nu$ so that $\dot{q}\rightarrow0$ and the region reaches thermal equilibrium with $P<u_{\rm B}$. In this case,  a static magnetosphere of trapped, hot matter is created. We discuss each possibility in turn, and then compare our estimates with the simulation data.

{\bf (1) Dynamical ejections:} Assuming that the density profile is fixed, the increase in $P$ from its initial value $P_i$ to $u_{\rm B}$ before ejection corresponds to an increase in the entropy of (eq.~\ref{s})
\beq
\frac{S_f}{S_i}\sim\left(\frac{u_{\rm B}}{P_i}\right)^{3/4}\,\,\,\,(u_{\rm B}>P_i)
\label{samp}
\eeq
where $S_i$ and $S_f$ are the initial and final entropy, respectively. Comparing equation (\ref{samp}) with Figure \ref{figure:pressure}, we see that for $B=5\times10^{15}$\,G we expect $S_f/S_i \sim 3$ in the trapped matter, and an increase in $\zeta\propto S^3$ of $\sim30$ (eq.~\ref{zeta_typical}), assuming that the material, once it escapes, does so with a dynamical expansion $t_{\rm dyn}$ timescale similar to the free unmagnetized wind. 

The characteristic timescale for $P$ to approach $u_{\rm B}$, and thus for the system to reach the critical condition for dynamical ejection, is 
\beq
t_{\rm ej}\sim\left(u_{\rm B}-P_i\right)/(\dot{q}\rho).
\eeq
For the purposes of an order-of-magnitude estimate of the trapping timescale $t_{\rm ej}$ and its scalings, we take $\dot{q}\sim H_\nu$ (eq.~\ref{h}) with $f(r)\simeq(1/2)(R_\nu/r)^2$ (eq.~\ref{fofr}), so $t_{\rm ej}\sim u_{\rm B}(1-P_i/u_{\rm B})/H_\nu\rho$, and we find that
\beq
t_{\rm ej}\sim0.05\,{\rm s}\frac{B_{16}^2 \,R^2_{\nu,10}(1-P_i/u_{\rm B})}{(L_{\bar{\nu}_e,\,51}/8)\langle\epsilon_{\bar{\nu}_e,\, \rm 14\,MeV}^2\rangle (X_{p}/0.45)\,\rho_8}\left(\frac{2R_\nu}{r}\right)^4 
\label{tej}
\eeq
where we have scaled to parameters appropriate to the high luminosity model shown in Figure \ref{figure:highl} --- $L_{\bar{\nu}_e,\,51}=8$, $X_{p}\simeq0.45$, $\langle\epsilon_{\bar{\nu}_e,\, \rm 14\,MeV}^2\rangle\simeq 410$\,MeV$^2$. Note that $t_{\rm ej}$ decreases strongly with radius so that the material nearest the PNS will be trapped the longest, so long as $u_{\rm B}$ is greater than the initial thermal pressure.

 {\bf (2) Static magnetosphere with $\dot{q}\rightarrow0$:} The second possibility is that the temperature increases such that $C_\nu\rightarrow H_\nu$ before the condition $P>u_{\rm B}$ is exceeded. If this happens, we assume the magnetosphere achieves dynamical equilibrium at a temperature specified by $H_\nu=C_\nu$, which corresponds to a temperature of
\beqa
T_{\star}&\simeq&3.1\,\,{\rm MeV}\left[\frac{X_p}{0.5}\frac{L_{\bar{\nu}_e,\,51}}{8}\frac{\langle\epsilon_{\bar{\nu}_e,\, \rm MeV}^2\rangle}{410}\frac{1}{R_{\nu,\,10}^{2}}\right]^{1/6} \nonumber \\
&\times&\left[1+\frac{X_n}{X_p}\frac{L_{\nu_e}}{L_{\bar{\nu}_e}}\frac{\langle\epsilon^2_{\nu_e,\,\rm MeV}\rangle}{\langle\epsilon^2_{\bar{\nu}_e\,\rm MeV}\rangle}\right]^{1/6}\left(\frac{2R_\nu}{r}\right)^{1/3},
\label{tstar}
\eeqa
where we have again assumed $f(r)\simeq(1/2)(R_\nu/r)^2$, and the second term in square brackets adds a correction of $\simeq1.07$ for $X_p/X_n=1$, $L_{\nu_e}/\lumanue=1/1.3$, and $\langle\epsilon^2_{\nu_e,\,\rm MeV}\rangle/\langle\epsilon^2_{\bar{\nu}_e\,\rm MeV}\rangle\simeq0.62$ ($\langle\varepsilon_{\nu_e}\rangle/\langle\varepsilon_{\rm \bar{\nu}_e}\rangle=11/14$. At fixed density, $T_\star$ corresponds to a critical entropy
\beq
S_\star\propto T_\star^3/\rho
\label{sstar}
\eeq
as in equation (\ref{s}). Thus, if the magnetic field is strong enough compared to the thermal pressure force, we expect the amplified entropy $S_f$ to reach at most $S_\star$ and not the estimate in equation (\ref{samp}). 

{\bf Model results:} In the right panel of Figure \ref{figure:pressure} we show an example calculation with $\lumanue=8\times10^{51}$\,ergs/s and $B=10^{16}$\,G, as in the dynamical calculations presented in Figures \ref{figure:highl}, \ref{figure:mass} (left panel), and \ref{figure:hz}. The solid red line shows the temperature $T$ and the dotted blue line shows $S_i$ in the free un-magnetized wind. From the left panel of  Figure \ref{figure:pressure}, comparing the top black solid line with the top dashed red line, we see that $R_\beta\simeq130$\,km (where $P=u_B$). The solid blue curve labeled $S_f$ shows the estimate from equation (\ref{samp}) and the dashed blue curve shows $S_\star$ for the same model. We see that $S_f>S_\star$ for $r<R_\star\simeq30$\,km. This means that the magnetic field is strong enough that we expect $T\rightarrow T_\star$ and $S\rightarrow S_\star$ for $r<R_\star$, and we therefore expect this region to be dynamically stable. However, for $R_\star < r<R_\beta$, $S_f<S_\star$. In this region we expect a large entropy enhancement given by equation (\ref{samp}) {\it and} we expect dynamical ejection on a timescale given by equation (\ref{tej}). 

Our estimate for the maximum entropy of the material ejected is then $S_{\rm max}=S_f(R_\star)\simeq700$, and the ejection timescale at $R_\star$ is of order $250$\,ms. Both are larger than the values we find in our simulations, but have strong dependencies on $B$ and $r$ over the range $R_\star<r<R_\beta$. Looking closely at the upper panels (e.g., upper left) of Figure \ref{figure:highl} we see that, along the equator, the entropy rapidly increases from $R_\nu$ to $\sim30$\,km, in agreement with the estimate of $R_\star$ in the right panel of Figure \ref{figure:pressure}. The radial extent of the entropy amplification seen upper left panel is also in fair agreement with $R_\beta\sim130$\,km predicted.

In the right panel of Figure \ref{figure:smax} we compare the maximum entropy we find in the simulations (black dots, lines) with the simple model described above, $S_{\rm max}=S_f(R_\star)$ (blue dotted), for a wide range of neutrino luminosity and magnetic field. For each neutrino luminosity, we take the $B=0$ hydrodynamical model and overlay a $u_B$ profile (eq.~\ref{tension}). We then calculate the location of $R_\star$ and $R_\beta$ in a manner analogous to described for the right panel of Figure \ref{figure:pressure}. We calculate the entropy enhancement with equation (\ref{samp}) in the range $R_\star \leq r \leq R_\beta$ and record the maximum value, $S_{\rm max}=S_f(R_\star)$. For the purposes of a first comparison, we take a constant value of the curvature radius of the magnetic field $R_c=R_\nu/2$ (eq.~\ref{tension})

The semi-analytic model described agrees at order-of-magnitude with the simulations. In particular, it does a reasonable job of estimating the magnetic field strength at which we see order-unity increases in the entropy. At low neutrino luminosity ($\lumanue=0.4\times10^{51}$\,ergs/s) the model significantly under-predicts $S_{\rm max}$, while at higher neutrino luminosity and high $B$ it over-predicts $S_{\rm max}$. Within the context of the model as described, these changes could be accommodated by a neutrino luminosity dependent radius of curvature $R_c(\lumanue)$. For example, taking $R_c=R_\nu$ in equation (\ref{tension}) gives a much better match to the simulation data for $\lumanue=8\times10^{51}$\,ergs/s, while $R_c=R_\nu/5$ gives a better match to  $\lumanue=0.4\times10^{51}$\,ergs/s. It is worth emphasizing again that the model neglects dynamical changes in the gas density $\rho$ as a function of radius, which directly impacts our estimate of $t_{\rm ej}$.

Lastly, we can use the model to predict how much matter has enhanced entropy ($R_\star < r<R_\beta$), is heated such that $\zeta>\zeta_{\rm crit}$, or is permanently trapped ($r<R_\star$) based on the fraction of the PNS surface corresponding to a given equatorial radial range, and assuming a dipole field geometry. For any given radius $r$ we define 
\beq
\delta(r)=1-\left(1-\frac{R_\nu}{r}\right)^{1/2}.
\eeq
Taking the model shown in the right panel of Figure \ref{figure:pressure} as an example, for $r=R_\star\simeq30$\,km, $\delta\simeq0.2$. If the region with $r<R_\star$ is in fact permanently trapped, this estimate implies that the total mass loss rate from the PNS would be decreased from the unmagnetized calculation by a factor of $\sim5$. Comparing our time-averaged values of the mass loss rate from the magnetized and unmagnetized simulations for this luminosity and magnetic field, we find a decrease of a factor of $\sim3$, in fair agreement with the model. Similarly, given the profile of $S_f(R_\star<r<R_\beta)$ and {\it assuming} that the matter eventually escapes with dynamical expansion timescale $t_{\rm dyn}$ equal to the value in the unmagnetized spherical wind calculation (see eq.~\ref{zetacrit}), we can estimate the total fraction of mass with $\zeta>\zeta_{\rm crit}$ in this region. Again referencing the right panel of Figure \ref{figure:pressure}, we estimate that a fraction $\sim0.1$ of the mass is ejected with $\zeta>\zeta_{\rm crit}$, which is in good agreement with our calculations from the simulations (see Section \ref{section:mass}). 

\section{Conclusions \& Discussion}
\label{section:discussion}

We present a first exploration of the dynamics and potential nucleosynthesis of neutrino-heated winds from proto-neutron stars born with strong dipole magnetic fields. Our results imply that magnetar-strength fields dominate the wind dynamics during the PNS cooling phase (Figs.~\ref{figure:highl}-\ref{figure:smax}). In particular, we find dramatically enhanced entropy in the dynamically-ejected plasmoids that emerge from the equatorial helmet streamer configuration. These eruptions are periodic at high neutrino luminosities (Fig.~\ref{figure:mass}), with a small amount of mass ejected in high-$\zeta$ material per eruption (Fig.~\ref{figure:hz}). In analogy with numerical studies of the Solar wind, as a result of magnetic confinement and continued heating, the structure cannot relax to a time-steady solution (see Section \ref{section:analytic}; \citealt{endeve2003,endeve2004}). As a result of the periodic ejections, the streamlines of the high-latitude free wind also undergo time-dependent modulation.  At low neutrino luminosities and lower magnetic field strengths, we find intermittent ejections (e.g., Figs.~\ref{figure:lowl}, \ref{figure:mass}). Once the magnetic field becomes dynamically dominant (see eq.~\ref{bcrit}), the entropy rapidly increases (see Fig.~\ref{figure:smax}). Our results are in qualitative agreement with the semi-analytic estimates of \cite{thompson03} (Section \ref{section:analytic}).

An important caveat is that in order to attain the high entropies reported here, the magnetospheric structure must be stable on the timescale of the plasmoid ejections, which range from $\sim50-500$\,ms. Although this timescale is short when compared to the PNS cooling time, it is long compared to both typical rotation periods expected for PNSs at birth and the PNS convective turnover timescale throughout the cooling epoch. Shear and convection may thus disrupt the large-scale magnetospheric structure needed for confinement and entropy amplification before $\zeta_{\rm crit}$ is exceeded. While this might lead to lower entropy in the extended magnetosphere where we find the ejections originate ($r\sim20-100$\,km in Fig.~\ref{figure:highl}, top panels), the more compact high entropy closed zone on the scale of $r\sim10-20$\,km may be sheared open, releasing matter with thermodynamic conditions markedly different than we find (e.g., Fig.~\ref{figure:hz}). Ejections precipitated by shearing motions and PNS convection might also significantly increase the total mass of outflowing material. Even in the absence of shearing and convection, rotation coupled with strong magnetic fields leads to magneto-centrifugal acceleration, which will change the entropy and dynamical timescale of the emerging matter (see below).

Based on surveys of nucleosynthesis in PNS winds (e.g., \citealt{hoffman97}; Section \ref{section:rprocess}), the thermodynamic conditions we find in the ejecta will produce the heavy $r$-process elements (eq.~\ref{zetacrit}) if the medium is neutron-rich. As found by previous studies, the high-latitude wind material is a guaranteed source of 1st- and possibly 2nd-peak $r$-process material, with a broad range of $\zeta$ for the wind material (see Fig.~\ref{figure:hz}).  Past explorations of $r$-process nucleosynthesis have explored a distribution of entropy, electron fraction, and dynamical timescale for PNS winds, attempting to constitute the observed Solar $r$-process abundances from different dynamical components. Looking at Figures \ref{figure:highl} and {\ref{figure:lowl} we see that an individual PNS may generate a very wide range of $S$ as a function of latitude. If the wind is strongly magnetized and neutron rich it will produce a broad distribution of abundances, with most of the material yielding a ``weak" $r$-process extending to the first abundance peak, less material producing the second peak, and even less extending to the third peak and beyond.

If the wind is instead proton-rich, the $\nu$p-process should operate and be strongly affected by the high entropy and short ejection timescales (Section \ref{section:pnuclei}).  We expect the mass number of the heaviest elements synthesized to increase with entropy based on the work of \cite{pruet2006}, who found that the maximum mass number of synthesized material increased from $A\simeq102$, to 120, to 170, as the entropy was artificially increased by a factor of 2 and 3, respectively. The large range of entropy we find in magnetized PNS winds thus suggests a broad range in maximum $A$ synthesized.

The high entropies and complicated dynamical expansion history of the matter ejected may produce a unique abundance distribution that can be probed with observations of stars formed from enriched ejecta, and former binary companions that are either now seen as high-velocity runaway stars, or are still within their companion's supernova remnant. As discussed in Section \ref{section:observations}, highly-magnetized neutron star birth may provide for early heavy metal enrichment in the history of the universe, observable pollution in binaries, and perhaps an explanation for the peculiar abundances of some stars.

Taken at face value, our calculations indicate that highly-magnetized neutron stars cannot be the dominant contributor to the $r$-process budget in the Galaxy. \cite{woods_thompson} argue that $\sim10$\% of all neutron stars are born as magnetars, eventually leading to the Anomalous X-Ray Pulsars and Soft Gamma-Ray Repeaters. Given a supernova rate in the Galaxy of $\Gamma_{\rm SN}\sim0.02$ yr$^{-1}$, the rate of magnetar production is of order $\sim2\times10^{-3}$ yr$^{-1}$. The production rate of heavy $r$-process elements with $A>130$ is of order $\sim10^{-7}$\,M$_\odot$/yr when averaged over the history of the Galaxy's star formation \citep{qian2000}. This would imply that {\it if} magnetars were to dominate the $r$-process production in the Galaxy they would need to produce $\sim5\times10^{-5}$\,M$_\odot$ of $r$-process material per birth. Yet, our estimates during the steady wind epoch in Section \ref{section:mass} imply that only $\sim10^{-5}$\,M$_\odot$ is ejected in high-$\zeta$ material. Given the $\alpha$-rich freezeouts expected from the preceding $\alpha$-process, we expect only $\sim10^{-6}$\,M$_\odot$ of heavy elements produced per highly-magnetized neutron star birth (depending on $Y_e$). This yield is sufficiently low that even if {\it all} neutron stars were born with transient short-lived high magnetic fields, they would still underproduce the claimed budget of Galaxy-averaged $A>130$ $r$-process nuclides. Unless we find unexpectedly that the threshold $\zeta$ for 3rd-peak $r$-process is somehow modified from that in \cite{hoffman97} in the non-standard thermal histories of the escaping high-$\zeta$ material, it seems that non-rotating highly-magnetized neutron stars cannot dominate $r$-process production. Because the mass loss rate increases with neutrino luminosity and PNS radius, a remaining option in the context of our models is that perhaps a single high-entropy ejection dominates the mass loss rate in $\zeta>\zeta_{\rm crit}$ material, just as the explosion commences, right at the start of the cooling epoch. Self-consistent calculations from the explosion to the wind phase are necessary to evaluate this possibility.

An additional piece of physics crucial to the dynamics of highly-magnetized PNS winds, but not included here, is rotation. Importantly, even a PNS rotation period of $P\sim3$\,ms, which implies a modest rotational energy reservoir of order $10^{51}$\,ergs, would give much higher expansion velocities of $\sim10^9-10^{10}$\,km/s on $50-100$\,km scales. We thus expect the dynamical expansion timescales to be significantly affected by the action of magneto-centrifugal acceleration \citep{tcq,metzger07,metzger08}. In static strong magnetic field configurations with rotation, \cite{vlasov,vlasov2017} find that magneto-centrifugal acceleration for $P\sim{\rm few}$\,ms can lead to factor of $\sim3-4$ increases in  $\zeta$ with respect to non-magnetic, non-rotating PNS winds. Combining rotation with the dynamical neutrino-heated magnetospheres we consider here is an important priority for future theoretical investigation. Rotation may also be important for the total yield of heavy elements from magnetized PNS birth. We note that, following models of super-luminous supernovae by \cite{kasen_bildsten} and \cite{woosley2010}, \cite{sukhbold_thompson} have recently shown as a proof of principle that modestly rotating magnetars with $P\sim3-5$\,ms and $B\sim10^{15}$\,G may produce normal Type IIP lightcurves. These same objects may produce unique heavy element nucleosynthesis via the dynamics reported here.

Many additional avenues for new work remain. The assumptions on the microphysics employed in this work should be relaxed (see Section \ref{section:numerical}). Specifically, the importance of Landau quantization of the electron-positron phase space should be assessed in dynamical calculations.  The flow should be laced with Lagrangian tracer particles for post-processing nucleosynthetic yields. In particular, the very rapid expansion we find just after plasmoid expulsion, may lead to interesting nucleosynthesis channels \citep{meyer2002,jordan_meyer}. In addition, some material in the figures has very long dynamical expansion timescale or can even have negative radial velocity after an eruptive event, potentially leading to long neutrino exposure times, perhaps heavy element nucleosynthesis, and then neutrino-induced spallation \citep{qian1997}. The numerical scheme employed should solve the equations of relativistic MHD to model the entire cooling epoch and the transition from non-relativistic thermally-driven wind to the Poynting-flux dominated pulsar-like phase \citep{komissarov_barkov}. More complex field topologies (e.g., quadrupole, octupole) and fully three-dimensional models should be explored. Following the evolution of the PNS radius, luminosity, and neutrino energies from explosion through the wind phase in General Relativity will lead to a better assessment of the total amount of material that can be ejected for a given dipole field strength. Perhaps an even more important outstanding issue as regards the numerical scheme is our use of ideal MHD. As a result, the magnetic reconnection events that occur in the dynamical ejections we see in our simulations are mediated only by uncontrolled numerical resistivity. 

\section*{Acknowledgments}

We thank Stan Woosley, Luke Roberts, and Brian Metzger for comments on the text. TAT thanks Brian Metzger, Eliot Quataert, Niccol\'o Bucciantini, Adam Burrows, and Jon Arons for conversations and collaboration on the $r$-process and magnetar winds over many years. TAT thanks Katra Byram for encouragement and support. AuD acknowledges support by NASA through Chandra Award numbers GO5-16005X, AR6-17002C, G06-17007B and TM7-18001X issued by the Chandra X-ray Observatory Center which is operated by the Smithsonian Astrophysical Observatory for and behalf of NASA under contract NAS8-03060.

\footnotesize{
\bibliographystyle{mnras}
\bibliography{references}
}

\end{document}